\documentclass[reqno,12pt]{article}
\usepackage{mdframed}
\usepackage{ae} 
\usepackage[T1]{fontenc}
\usepackage[ansinew]{inputenc}
\usepackage{amsmath}
\usepackage{amssymb}
\usepackage{graphicx}

\usepackage{color}
\definecolor{darkblue}{cmyk}{0.9,0.9,0,0}
\usepackage[colorlinks=true,linkcolor=darkblue,citecolor=darkblue,urlcolor=darkblue]{hyperref}
\usepackage{epsfig}

\usepackage[dvipsnames]{xcolor}
\usepackage{graphicx}

\newcommand{\beq}{\begin{equation}}
\newcommand{\eeq}{\end{equation}}
\newcommand\beqa{\begin{eqnarray}}
\newcommand\eeqa{\end{eqnarray}}
\newcommand\bea{\begin{array}}
\newcommand\eea{\end{array}}

\def\XXint#1#2#3{{\setbox0=\hbox{$#1{#2#3}{\int}$}
\vcenter{\hbox{$#2#3$}}\kern-.5\wd0}}

\newcommand{\nn}{\nonumber}

\newcommand{\neqa}{\nonumber\end{eqnarray}}
\newcommand{\la}[1]{\label{#1}}

\renewcommand{\d}{\partial}

\newcommand{\<}{{\langle}}
\renewcommand{\>}{{\rangle}}

\newcommand{\re}{\relax{\rm I\kern-.18em R}}

\renewcommand{\sp}{p\hspace{-.40em}/}

\def\su2{{SU(2)}}

\def\[{\left[}
\def\]{\right]}

\def\({\left(}
\def\){\right)}
\def\[{\left[}
\def\]{\right]}

\def\<{\langle}
\def\>{\rangle}

\def\i2{\frac{i}{2}}

\def\spi{\relax{\rm \pi\kern-0.5em /}}
\def\sA{\relax{\rm A\kern-0.5em /}}
\def\sp{\relax{\rm p\kern-0.5em /}}
\def\sd{\relax{\rm \d\kern-0.5em /}}
\def\sk{\relax{\rm k\kern-0.5em /}}
\def\sn{\relax{\rm n\kern-0.5em /}}
\def\sl{\relax{\rm l\kern-0.5em /}}
\def\sP{\relax{\rm P\kern-0.7em /}}
\def\sBethe{\relax{\rm \Bethe\kern-0.5em /}}

\newcommand{\nocontentsline}[3]{}
\newcommand{\tocless}[2]{\bgroup\let\addcontentsline=\nocontentsline#1{#2}\egroup}

\usepackage{varioref}
\usepackage{makeidx}
\makeindex

\newcommand\blfootnote[1]{%
  \begingroup
  \renewcommand\thefootnote{}\footnote{\hspace{-6mm}#1}%
  \addtocounter{footnote}{-1}%
  \endgroup
}

\usepackage[english]{babel}
\begin{document}


\thispagestyle{empty}

\renewcommand{\thefootnote}{\fnsymbol{footnote}}
\setcounter{page}{1}
\setcounter{footnote}{0}
\setcounter{figure}{0}

\vspace{-0.4in}

\begin{center}
$$$$
{\Large\textbf{\mathversion{bold}
 The Wilson Loop -- Large Spin OPE Dictionary 
}\par}
\vspace{1.0cm}

\textrm{Carlos Bercini$^\text{\tiny 1}$, Vasco Gon\c{c}alves$^\text{\tiny 1,\tiny 2}$, Alexandre Homrich$^\text{\tiny 1, \tiny 3, \tiny 4}$ Pedro Vieira$^\text{\tiny 1,\tiny 3}$}
\blfootnote{\tt  \#@gmail.com\&/@\{carlos.bercini,vasco.dfg,alexandre.homrich,pedrogvieira\}}
\\ \vspace{1.2cm}
\footnotesize{\textit{
$^\text{\tiny 1}$ICTP South American Institute for Fundamental Research, IFT-UNESP, S\~ao Paulo, SP Brazil 01440-070 \\
$^\text{\tiny 2}$Centro de Fisica do Porto e Departamento de Fisica e Astronomia, Faculdade de Ciencias da Universidade do Porto, Porto 4169-007, Portugal   \\
$^\text{\tiny 3}$Perimeter Institute for Theoretical Physics,
Waterloo, Ontario N2L 2Y5, Canada \\
$^\text{\tiny 4}$Walter Burke Institute for Theoretical Physics, California Institute of Technology, Pasadena, California 91125, USA
}  
\vspace{4mm}
}
\end{center}

\par\vspace{1.5cm}


\vspace{2mm}
\begin{abstract}
We work out the map between null polygonal hexagonal Wilson loops and spinning three point functions in large $N$ conformal gauge theories by mapping the variables describing the two different physical quantities and by working out the precise normalization factors entering this duality. By fixing all the kinematics we open the ground for a precise study of the dynamics underlying these dualities -- most notably through integrability in the case of planar maximally supersymmetric Yang-Mills theory.

\end{abstract}

\newpage

\setcounter{page}{1}
\renewcommand{\thefootnote}{\arabic{footnote}}
\setcounter{footnote}{0}



{
\tableofcontents
}



\newpage

\section{Introduction} 
In appendix B of~\cite{nullOPE} a duality was proposed between the $n$-point correlation functions of large spin single trace twist-two operators in planar $\mathcal{N}=4$ SYM and the expectation value of null polygonal Wilson loops with $2n$ sides.\footnote{This duality is one branch out of a rich web of dualities relating various seemingly distinct quantities in $\mathcal{N}=4$ SYM such as Wilson Loops and Scattering Amplitudes \cite{d1} and Wilson loop and the null limit of correlation functions \cite{Alday:2010zy}. Indeed, null correlation functions are dominated by leading twist large spin operators which is one way to argue for the duality mentioned in the main text. The argument in \cite{nullOPE} also uses some string theory intuition coming from the behavior of minimal surfaces of spinning strings and how they are expected to become related to the minimal surface describing null polygonal Wilson loops when their spin is taken to infinity. The argument was qualitative and no precise equality was spelled out in \cite{nullOPE}. Our main result (\ref{finalresult}) fills in this gap.} The simplest non-trivial example of such duality would relate three point functions and the null hexagon Wilson loop
\beq
\< \mathcal{O}_{J_1}(x_1,\epsilon_1) \mathcal{O}_{J_2}(x_2,\epsilon_2) \mathcal{O}_{J_3}(x_3,\epsilon_3) \> 
\longleftrightarrow
\mathbb{W}(U_1,U_2,U_3)  \la{relationGoal}
\eeq
The goal of this paper is to sharpen the arrow in this relation making it into a precise equation with an equal sign with all the appropriate normalizations and with a precise dictionary relating the variables on both sides of this equation: the spins $J_j$ and polarization vector $\epsilon_j$ on the left hand side and the hexagon cross-ratios $U_i$ on the right hand side. 

This is (\ref{finalresult}).\footnote{The reader might be frowning. In (\ref{finalresult}) there are $\ell_i$'s instead of $\epsilon_i$'s. Worry not, they are simply conjugate variables as reviewed in the next section and it is straightforward to change from one to the other. The map of kinematics using the spinors is given in (\ref{epsilonMap}).} 

We got there in two steps. First we examined the OPE decomposition of six point functions
 in the so-called snowflake channel: we fuse adjacent pairs of external operators into spinning operators which are then glued together through a tensor structure parametrized by integer indices $\ell_i$. The starting point is intimidating. It is given by 9 sums ($3$ are spin sums, $3$ are sums over tensor structures indices and the last $3$ appear in the representation of the relevant conformal block). When the external points approach the cusps of a null hexagon, six of these sums can be performed by saddle point. The location of the saddle point will fix the tensor structure indices $\ell_j$ to precise locations depending on the cross-ratios $U_j$ of the null hexagon. This gives us the map $\ell(U)$ spelled out in equation (\ref{lUmap}) below. Next 
%
 we analysed further the null six point correlator through an analytic bootstrap perspective (generalizing~\cite{Multi} -- where this was carried over for small $U_j$ in the so called origin limit~\cite{origin} -- to generic finite cross-ratios $U_j$). 
This allowed us to see how the correlators becomes Wilson loops and what are all the precise conversion factors showing up along the way. 
 
Null hexagon Wilson loops have light-cone singularities when non-adjacent vertices become null. We conjecture how these singularities \textit{emerge} from the discrete structure of the structure constants in the large spin limit. The limit (\ref{relationGoal}) should be understood to hold \textit{before} light-cones are crossed, i.e. in the ``Euclidean" region of positive cross ratios. Configurations with time-like separations should then be achieved through analytic continuation from this safe region. These musings are backed up by explorations of novel one loop data we extract.

In sum, in this paper we cleaned up the kinematics behind the duality (\ref{relationGoal}) using bootstrap techniques. We are currently investigating its dynamics with integrability.

\section{Spinning Three Point Functions} \la{reviewSec}
The purpose of this section is to establish notation. A traceless symmetric, spin $J$, primary operator in a CFT can be represented through an homogenous polynomial of degree $J$ on an auxiliary null polarization vector $\epsilon$
\begin{equation}
\mathcal{O}_J(x,\epsilon) = \epsilon_{\mu_1} \dots  \epsilon_{\mu_J}  \mathcal{O}^{\mu_1 \dots \mu_J}(x).
\end{equation}
In a parity preserving 4D CFT, three point functions of traceless symmetric parity even operators can be parametrized as 
\begin{equation}
\langle \mathcal{O}_{J_1} (x_1, \epsilon_1), \mathcal{O}_{J_2} (x_2, \epsilon_2) , \mathcal{O}_{J_3} (x_3, \epsilon_3) \rangle =\frac{ \sum\limits_{\ell_i } C^{J_1,J_2,J_3}_{\ell_1, \ell_2,\ell_3} {V_{1,23}^{J_1 - \ell_2 - \ell_3} }{V_{2,31}^{J_2 - \ell_3 - \ell_1}} {V_{3,12}^{J_3 - \ell_1 - \ell_2}} {H_{23}^{\ell_1}}  {H_{31}^{\ell_2}}  {H_{12}^{\ell_3}}}{(x_{12}^2)^{\frac{\kappa_1 + \kappa_2 - \kappa_3}{2}}(x_{23}^2)^{\frac{\kappa_2 + \kappa_3 - \kappa_1}{2}}(x_{31}^2)^{\frac{\kappa_3 + \kappa_1 -\kappa_2}{2}}}, \label{3pt}
\end{equation} 
where  $\kappa_i$ is the conformal spin and  
\begin{equation} V_{i,jk} =\left(\epsilon_i \cdot  x_{ik} x_{ij}^2 - \epsilon_i \cdot  x_{ij} x_{ik}^2\right)\frac{1}{x^2_{jk}},  \qquad \nonumber H_{ij} = \epsilon_i \cdot x_{ij} \epsilon_j \cdot x_{ij} - \tfrac{1}{2} x^2_{ij} \epsilon_i \cdot \epsilon_j,
\end{equation}
are a basis of conformal covariant tensors \cite{Costa:2011mg}, see appendix \ref{app:spinnors}. We sum over all non-negative integers $\ell$'s such that all exponents in (\ref{3pt}) are non-negative. 

Henceforth we will consider twist two operators in planar $\mathcal{N}=4$ SYM at weak coupling and use the short-hand notation ${C}^{\bullet  \bullet \bullet} \equiv C^{J_1,J_2,J_3}_{l_1, l_2,l_3}$ for the structure constants of three spinning operators. We also have 
\beq
{C}^{\bullet  \bullet \bullet}  = \hat{C}^{\bullet  \bullet \bullet} \times \underbrace{  \prod_{i=1}^3 \frac{J_i!^2}{(\ell_i!)^2\sqrt{(2J_i)!}(J_i+\ell_i-\sum_{j=1}^3 \ell_j)!}  }_{{C}^{\bullet  \bullet \bullet}_\text{tree level}}
\eeq
where $\hat{C}^{\bullet   \bullet \bullet}\equiv {C}^{\bullet  \bullet \bullet}/{C}_\text{tree level}^{\bullet  \bullet \bullet}$ is given by an expansion in small 't Hooft coupling $\lambda$ and captures all loop corrections. 

\section{Null Correlators and the $U(\ell)$ map} \la{OPEG6Sec} 

We consider the null polygonal limit of the six point correlator of the lightest single trace gauge invariant scalar operators as in \cite{Multi}. This correlator is given by $9$ cross-ratios carefully reviewed in appendix \ref{appCR}. We will sequentially send $6$ of them to zero when taking each $x_i$ to be null separated from $x_{i+1}$ to obtain in the end a function which depends on the remaining $3$ cross-ratios. More precisely, the final result will depend on the three finite cross-ratios as well as logs of the six vanishing cross-ratios. The dependence on the latter will be through a factorized universal pre-factor which we can fix. The dependence on the finite cross-ratios will be related to the renormalized Wilson loop which is theory dependent.

\begin{figure}[t]
\includegraphics[width=\textwidth]{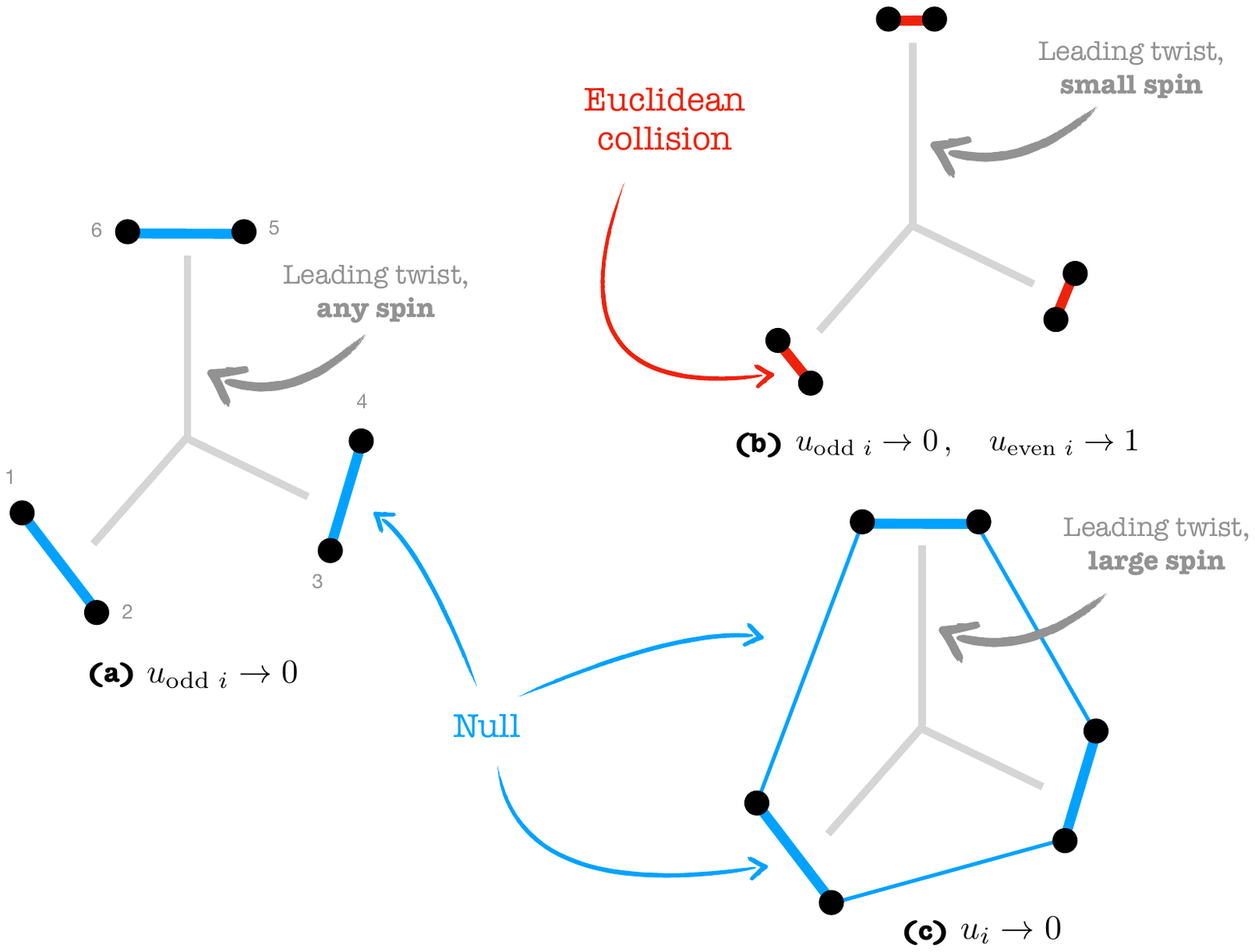}
\caption{\small Various snow-flake OPE limits discussed in this paper. The bottom right one is the double light-like OPE explored in this section. The top right one is the more conventional Euclidian OPE used in appendix \ref{app:Casimirs} to extract new one loop OPE data which is analysed in appendix \ref{appendix1Loop}. We can get to both starting from the single light-like OPE on the left.}
\label{snow}
\end{figure}

As explained in \cite{Multi} we can project into leading twist (i.e. two) in the $12$, $34$ and $56$ channel in the snowflake decomposition by taking $u_1,u_3,u_5\to 0$ or $x_{12}^2,x_{34}^2,x_{56}^2\to 0$ as depicted in figure \ref{snow}a. In this limit, in perturbation theory the six point function behaves as 
\beq
G_6({\color{red}u_1},u_2,{\color{red}u_3},u_4,{\color{red}u_5},u_6,U_1,U_2,U_3) \to {\color{red} u_1 u_3 u_5}\,  \hat{G}_6(u_2,u_4,u_6,U_1,U_2,U_3) 
\eeq
The function $\hat{G}_6$ has no powers of $u_1, u_3$ or $u_5$ but it {{implicitly}} contains arbitrarily many powers of $\ln(u_1),\ln(u_3)$ and $\ln(u_5)$ arising from the anomalous dimensions of the twist two operators. This function can be expanded as 
 \beq
G_6 = \sum_{J_1,J_2,J_3} \sum_{\ell_1,\ell_2,\ell_3}  \underbrace{\hat P^{\bullet \bullet \bullet}(J_1,J_2,J_3,\ell_1,\ell_2,\ell_3)}_{\texttt{dynamic}} \int\limits_{y_j\in[0,1]}  dy_1 dy_2 dy_3  \underbrace{\mathcal{F}(J_i,\ell_i,y_i,u_i,U_i)  }_\texttt{kinematics}\label{start} 
\eeq
where $\hat P$ is a (theory dependent) normalized product of three point functions\footnote{It is given by a product of three three point functions of two scalar and one spinning operator (for the three OPE's of the $12$, $34$ and $56$ OPE's of the external scalar operators) and a fully spinning three point function (the intersection of the three gray lines in figure \ref{snow}), 
\beq
\hat P^{\bullet \bullet \bullet}(J_1,J_2,J_3,\ell_1,\ell_2,\ell_3)=\hat C^{\bullet \bullet \bullet}(J_1,J_2,J_3,\ell_1,\ell_2,\ell_3) \hat C^{\circ \circ \bullet}(J_1)\hat C^{\circ \circ \bullet}(J_2)\hat C^{\circ \circ \bullet}(J_3) \,. \la{PC}
\eeq
Here the hat $\hat C$ stands for tree-level normalized quantity $C/C_\text{tree level}$. 
} and ${\mathcal{F}}$ is a (theory independent) conformal block integrand worked out in \cite{Multi} and recalled in appendix~\ref{Fappendix}. 

Series expanding the left and right hand side of relation (\ref{start}) around $u_i,U_i=1$ -- corresponding to the conventional Euclidean OPE limit depicted in figure \ref{snow}b -- allows us to extract structure constants $\hat P$ for the lowest spins $J$'s and polarization integers $\ell$'s. This data extraction using the one loop result \cite{Drukker:2008pi} for $G_6$ is described in appendices \ref{app:Casimirs}, \ref{appendix1Loop}. This one loop OPE data will be used in section 5.

In this section, we consider instead the limit $u_2,u_4,u_6\to0$ (at fixed $U_j$) -- {{known as the Lorentzian null OPE limit}} depicted in figure \ref{snow}c -- which is realized when all external points approach the cusps of a null hexagon which in turn is parametrized by the finite cross-ratios $U_i$. In this limit
\beq
\hat{G}_6(u_2,u_4,u_6,U_1,U_2,U_3) \to {\color{red} u_2u_4 u_6}\,\widetilde{G}_6(U_1,U_2,U_3) 
\eeq
where $\widetilde{G}$ is a non-trivial function of the finite $U_i$ which still contains arbitrary powers of $\ln(u_j)$ but no powers of $u_j$ since these were by now all sent to zero. 
We find two important simplifications when computing the correlator (\ref{start}) in this $u_j \to0 $ limit:
\begin{itemize}
\item The integral is dominated by large spins $J_j$ and large polarization integers $\ell_j$. We can thus transform $$\sum\limits_{J_1,J_2,J_3} \sum\limits_{\ell_1,\ell_2,\ell_3}  \to \frac{1}{8} \int\limits_{0}^\infty dJ_1 dJ_2 dJ_3 d\ell_1 d\ell_2 d\ell_3$$ in (\ref{start}) being left with nine integrals in total. (The $8=2^3$ comes from the fact that the spins $J$ are even.)
\item Six of those nine integrals can be done by saddle point.
\end{itemize}
More precisely, we find that $0=\partial \ln \mathcal{F} /\partial \ell_j =\partial \ln \mathcal{F} /\partial y_j $ 
leads to the saddle point location   
\beqa
   y_1= \frac{J_2 }{J_2 +J_3 \sqrt{\color{blue} \frac{U_2 U_3}{U_1}}} \, , \qquad   y_2 = \frac{J_3 }{J_3 +J_1 \sqrt{\color{blue} \frac{U_1 U_2}{U_3}}}\, ,\qquad    y_3= \frac{J_1 }{J_1+J_2   \sqrt{\color{blue} \frac{U_1 U_3}{U_2}}}
\eeqa
and more importantly 
\beqa
   {\color{magenta}\ell_1} = \frac{J_2 J_3}{J_2+J_3+J_1\sqrt{\color{blue} \frac{U_2}{U_1 U_3}}}\, , \qquad      {\color{magenta}\ell_2}= \frac{J_1 J_3}{J_1+J_3+J_2\sqrt{\color{blue} \frac{U_1}{U_2 U_3}} }\,, \qquad      {\color{magenta}\ell_3}= \frac{J_1 J_2}{J_1+J_2+J_3\sqrt{\color{blue} \frac{U_3}{U_1 U_2}}} \label{lUmap} \, 
\eeqa
which nicely relate the Wilson loop cross-ratios in the right hand side of (\ref{relationGoal}) with the spin and polarization integers appearing in structure constant in the left hand side of this relation. They are the sought after dictionary between these two worlds. (If $J_i \gg \ell_i \gg 1$ then the $U_i$ are very small; this was the limit studied in \cite{Multi}.)

The saddle point evaluation leads to
\begin{align}
\widetilde{G}_6 = & \frac{4 \color{red}{u_2 u_4 u_6}}{({\color{blue}U_1 U_2 U_3})^{\frac{1}{2}}}\int_0^{\infty} dJ_1\,dJ_2\,dJ_3 \,\hat{P}^{\bullet\bullet\bullet}(J_1,J_2,J_3,\ell_1,\ell_2,\ell_3) e^{
-\frac{J_1 J_2}{J_3}\times  \frac{{\color{red} u_2}\sqrt{{\color{blue} U_2}}}{\sqrt{\color{blue} U_1 U_3}}
-\frac{J_2 J_3}{J_1}\times  \frac{{\color{red} u_4}\sqrt{\color{blue} U_1}}{\sqrt{\color{blue} U_2 U_3}}
-\frac{J_1 J_3}{J_2}\times  \frac{{\color{red} u_6}\sqrt{\color{blue} U_3}}{\sqrt{\color{blue} U_1 U_2}}} \nonumber \\
& \times 2^{\gamma_1+\gamma_2+\gamma_3}\left(\frac{{\color{red}u_1}}{J_1}\right)^{\frac{\gamma_1}{2}}
\left(\frac{{\color{red}u_3}}{J_2}\right)^{\frac{\gamma_2}{2}}
\left(\frac{{\color{red}u_5}}{J_3}\right)^{\frac{\gamma_3}{2}}\left(\frac{\ell_1}{{\color{blue}U_1}^\frac{1}{2}}\right)^{\frac{-\gamma_1+\gamma_2+\gamma_3}{2}}\left(\frac{\ell_2}{{\color{blue}U_3}^\frac{1}{2}}\right)^{\frac{\gamma_1-\gamma_2+\gamma_3}{2}}\left(\frac{\ell_3}{{\color{blue}U_2}^\frac{1}{2}}\right)^{\frac{\gamma_1+\gamma_2-\gamma_3}{2}},
\label{G6Saddle}
\end{align}
where $\ell_j$ depend on the integration variables $J_j$ through (\ref{lUmap}). Implicit in this discussion is the assumption that the integral is dominated by the saddle point developed by the conformal block integrand. This should be valid for positive $U$s, see further discussion in section \ref{oneloop}. One can nicely check that when $\lambda=0$ (so that the full second line as well as $\hat{P}^{\bullet\bullet\bullet}$ can be set to 1) this expression indeed integrates into the free theory result $\widetilde{G}_6 = 1$.

We close this section with the inverse of the map (\ref{lUmap}):
\beqa
&&{\color{blue} U_1}=
\frac{J_1 J_3   {\color{magenta} \ell_1   \ell_3}}{\left(\left(J_2+J_3\right)   {\color{magenta} \ell_1}-J_2 J_3\right)
   \left(\left(J_1+J_2\right)  {\color{magenta}  \ell_3}-J_1 J_2\right)} \nn \,,\\
   &&{\color{blue} U_2}=\frac{J_2 J_3   {\color{magenta} \ell_2   \ell_3} }{\left(\left(J_1+J_3\right)   {\color{magenta} \ell_2   }-J_1 J_3\right) \left(\left(J_1+J_2\right)   {\color{magenta} \ell_3}-J_1
   J_2\right)} \,, \nn \\ 
   && {\color{blue} U_3}=\frac{J_1 J_2   {\color{magenta} \ell_1   \ell_2}}{\left(\left(J_2+J_3\right)     {\color{magenta} \ell_1   }-J_2 J_3\right)
   \left(\left(J_1+J_3\right)     {\color{magenta} \ell_2}-J_1 J_3\right)} \, \label{Ulmap}.
\eeqa
It is going to be used intensively below.
\section{Multi-point Null Bootstrap and the $C_{123}/\mathbb{W}$ relation} 
\la{BootstrapSection}
We took a limit where all points approach the boundary of a null hexagon corresponding to all $u_j \to 0$. Because we did it in two steps (first $u_1,u_3,u_5 \to0$ projecting to leading twist and then $u_2, u_4, u_6 \to0$ projecting to large spin) the final result (\ref{G6Saddle}) is not manifestly cyclic invariant. In this section we follow \cite{Multi} and impose the cyclic symmetry of our correlator under $u_i \to u_{i+1}$ and $U_i \to U_{i+1}$ to further constraint the structure constants $\hat P$. This will generalize the result in \cite{Multi} from the origin kinematics to generic hexagon cross-ratios.

To kick this analysis off we start by converting the starting point (\ref{G6Saddle}) from the cross-ratios $u_j$ to the more local cross-ratios $v_j$ (both are reviewed in appendix \ref{appCR}) since the expectation is that the Wilson loop should factorize into a universal prefactor depending on these variables alone times a renormalized Wilson loop \cite{Alday:2010zy,Multi}. Beautifully, we see that this factorization is almost automatic once we convert to the $v$ variables. Indeed, we find
\begin{align}
\widetilde{G}_6 = &\,4\, \sqrt{v_2v_4v_6}  \hspace{2pt} \!\int\limits_0^{\infty}\! dJ_1\,dJ_2\,dJ_3 \,  e^{
-\frac{J_1 J_2}{J_3}\,\sqrt{\frac{v_2 v_6}{v_4}}
-\frac{J_2 J_3}{J_1}\,\sqrt{\frac{v_2 v_4}{v_6}}
-\frac{J_1 J_3}{J_2}\,\sqrt{\frac{v_4 v_6}{v_2}}+ \frac{\gamma_1 }{4} \ln \frac{16 v_1 v_5}{v_3 J_1^2} +\frac{\gamma_2 }{4} \ln \frac{16 v_1 v_3}{v_5 J_2^2} +\frac{\gamma_3 }{4} \ln \frac{16 v_3 v_5}{v_1 J_3^2}  } \nn
\\&\,\times \hat{P}^{\bullet\bullet\bullet}(J_1,J_2,J_3,\ell_1,\ell_2,\ell_3)/\prod_{i=1}^3 \ell_i^{\frac{\gamma_i-\gamma_{i+1}-\gamma_{i-1}}{2}} \label{G6v}  
\end{align}
so that the first line is already only made out of $v_j$'s while all $U_j$ dependence arises from the second line through the $\ell_j(J_i,U_i)$ map (\ref{lUmap}). The problem at this point is how to constrain $\hat P$ so that the $U_j$ and $v_j$ dependence factorizes and so that the final result is cyclic invariant under $v_j, U_j \to v_{j+1},U_{j+1}$. The factorization would be automatic as soon as the  $\ell_j$ dependence in~$\hat P$ comes through a factor of the form 
\beqa
&&\texttt{factor} \equiv \prod_{i=1}^3 {\color{magenta} \ell_i}^{\frac{\gamma_i-\gamma_{i+1}-\gamma_{i-1}}{2}}\, \times\\
&& \times \mathbb{W}(\tfrac{J_1 J_3   {\color{magenta} \ell_1   \ell_3}}{\left(\left(J_2+J_3\right)   {\color{magenta} \ell_1}-J_2 J_3\right)
   \left(\left(J_1+J_2\right)  {\color{magenta}  \ell_3}-J_1 J_2\right)} , \tfrac{J_1 J_3   {\color{magenta} \ell_1   \ell_3}}{\left(\left(J_2+J_3\right)   {\color{magenta} \ell_1}-J_2 J_3\right)
   \left(\left(J_1+J_2\right)  {\color{magenta}  \ell_3}-J_1 J_2\right)},\tfrac{J_1 J_3   {\color{magenta} \ell_1   \ell_3}}{\left(\left(J_2+J_3\right)   {\color{magenta} \ell_1}-J_2 J_3\right)
   \left(\left(J_1+J_2\right)  {\color{magenta}  \ell_3}-J_1 J_2\right)}) \, \nn
\eeqa
Indeed, the first factor would cancel precisely the factor in the denominator in the last line of~(\ref{G6v}) whereas -- on the saddle point solution (\ref{lUmap}) -- the arguments of the second function will become precise the $U_j$ variables as indicated in (\ref{Ulmap}). That is, if 
\begin{equation}
\hat P^{\bullet \bullet \bullet}(J_1,J_2,J_3, \ell_1, \ell_2, \ell_3)= \texttt{factor} \times p(J_1,J_2,J_3)\,.
\label{PhatAnsatz}
\end{equation}
then we automatically find an explicit factorization
\begin{align}
\widetilde{G}_6 = &\mathbb{W}(U_1,U_2,U_3) \times \Big[ \,4 \sqrt{v_2v_4v_6}  \hspace{2pt} \!\int\limits_0^{\infty}\! dJ_1\,dJ_2\,dJ_3 \,  e^{
-\frac{J_1 J_2}{J_3}\,\sqrt{\frac{v_2 v_6}{v_4}}
-\frac{J_2 J_3}{J_1}\,\sqrt{\frac{v_2 v_4}{v_6}}
-\frac{J_1 J_3}{J_2}\,\sqrt{\frac{v_4 v_6}{v_2}}} \label{G6v2}  \\
&\!\!\!\!\! e^{\frac{f}{4} \ln(J_1) \ln\! \big(\!\frac{  v_1 v_5}{v_3}\!\big) +\frac{f}{4} \ln(J_2) \ln\!\big( \!\frac{  v_1 v_3}{v_5}\! \big)+\frac{f}{4} \ln(J_3) \ln\!\big(\! \frac{  v_3 v_5}{v_1}\! \big) +\frac{g}{4} \ln\!\big( \!\frac{16^3 v_1 v_3 v_5}{J_1^2J_2^2 J_3^2}\! \big)-\frac{f}{2} \sum\limits_{j=1}^3 \ln(J_j) \ln(J_j/4)  } p(J_1,J_2,J_3)   \Big] \nn
\end{align}
where we have used the explicit form of the large spin anomalous dimension $\gamma_i = f \ln(J_i) + g$ to massage the second line. 
It is hard to imagine how anything else would lead to a factorization but we did not establish the uniqueness of (\ref{PhatAnsatz}); it is a conjecture which passes some non-trivial checks below and reduces to \cite{Multi} in the origin limit.\footnote{The challenge is to relate (non-)factorization of integrands versus (non-)factorization of integrated expressions. Any extra $\ell_j$ dependence in (\ref{PhatAnsatz}) would show up inside the square bracket in (\ref{G6v2}) and thus generically lead to a $U_j$ dependence once we integrate in $J_j$ with (\ref{lUmap}). It might be that a very subtle $\ell_j$ dependence could integrate to zero or generate a factorized function of $U_j$  which would renormalize $\mathbb{W}$. We were not imaginative enough to find any such example which made us confident that (\ref{PhatAnsatz}) is indeed unique.}  

Next we have to impose cyclicity. For the first factor in (\ref{G6v2}) this simply means that $\mathbb{W}(U_1,U_2,U_3) = \mathbb{W}(U_2,U_3,U_1)$ but it does not constraint $\mathbb{W}$ any further. On the contrary, for the second factor, cyclicity is very powerful. It fixes $p$ completely to all loop orders in perturbation theory, under very mild assumptions as explained below. The result is remarkably simple:
\beq
p(J_1,J_2,J_3)= \mathcal{N}  \prod_{i=1}^3\left( \Gamma\left(1-\gamma_i\right) e^{\frac{f}{2}\ln(J_i)^2-f\ln 2 \ln J_i}\right) \label{finalresultppp}.
\eeq

It is a nice and very instructive exercise to plug this proposal into (\ref{G6v2}), expand the integrand to any desired order in perturbation theory (corresponding to small cusp anomalous dimension $f$ and small collinear anomalous dimension $g$), perform all the resulting integrations and realize that we only generate $\ln(v_j)$'s and that moreover the result non-trivially combines, order by order in perturbation theory, into a fully cyclic expression. 

It is an even more instructive exercise to simply plug a general perturbative ansatz for $p(J_1,J_2,J_3)$ as an infinite series of monomials made out of powers of $\ln(J_i)$'s in (\ref{G6v2}). Each such monomial will again integrate to simple polynomials in $\ln(v_j)$'s. Remarkably, imposing cyclicity at each order of perturbation theory will completely fix these polynomials and thus the full perturbative expansion up to an overall normalization constant. In this way, by considering a very large number of loops we could eventually recognize a simple pattern and arrive at (\ref{finalresultppp}). This brute force derivation is perfectly valid but was \textit{not} how we originally arrived at (\ref{finalresultppp}). 

We proceeded in a slightly more sophisticated way following similar ideas in the four point function analysis in \cite{Alday:2016mxe}. This is explained in the box that follows; this discussion can be probably skipped in a first reading. 

     \begin{mdframed}[frametitle={Deriving (\ref{finalresultppp})},frametitlealignment=\centering,backgroundcolor=red!6, leftmargin=0cm, rightmargin=0cm, topline=false,
	bottomline=false, leftline=false, rightline=false] 
\footnotesize 
We first look for \textit{an} integral transform for $p$ such that ciclicity can be imposed at integrand level. We define
 \beq
q\,(\ln J_1, \ln  J_2,\ln  J_3)=e^{-\frac{f}{2} \sum\limits_{j=1}^3 \ln(J_j) \ln(J_j/4)  } p(J_1,J_2,J_3) \label{qp}
\eeq to absorb the last factor in the exponential in the second line in (\ref{G6v2}) and we change integration variables to 
\begin{equation}
x_1=\frac{J_2 J_3}{J_1}\sqrt{\frac{v_2 v_4}{v_6}}, \qquad 
x_3= \frac{J_1 J_2}{J_3}\sqrt{\frac{v_2 v_6}{v_4}}, \qquad 
x_5 =\frac{J_1 J_3}{J_2}\sqrt{\frac{v_4 v_6}{v_2}}, \qquad 
\end{equation}  
to trivialize the tree level measure. Then the previous expression (\ref{G6v2}) takes the very suggestive form
\begin{align}
\widetilde{G}_6 = &\mathbb{W}(U_1,U_2,U_3) \times e^{ \sum\limits_{i} \frac{f}{16}\ln v_i \ln v_{i+3} -\frac{f}{8}\ln v_i \ln v_{i+1}+\frac{g}{4} \ln v_i}  \times \label{G6v3}  \\
&\times  \int\limits_0^{\infty}  dx_1\,dx_3\,dx_5 \,  e^{  - \sum\limits_{i=1}^3 (x_{2i-1}+ \frac{g}{2} \ln(x_{2i-1}/4)- \frac{f}{4} \ln(x_{2i-1}) \ln(v_{2i-1}))} \nn \\
&\qquad \qquad \times  q\(\frac{\ln x_1+\ln x_5-\ln v_6}{2},\frac{\ln x_3+\ln x_1-\ln v_2}{2},\frac{\ln x_3+\ln x_5-\ln v_4}{2}\) \nn 
\end{align}
where we see the explicit appearance of the Sudakov factor \cite{Alday:2010zy,Multi} in the first line. The lack of ciclicity is now quite striking in the very different way that the even and odd  cross-ratios show up in the integral: The odd cross-ratios appear in the exponent in the form $\ln(x_i) \ln(v_i)$ while the even cross-ratios appear inside the arguments of the dressed structure constant $\hat p$. That asymmetry is trivial to fix: It suffices to write $q$ itself as an integral transform introducing three new integration variables $x_2,x_4,x_6$ as 
\begin{align}
q(X,Y,Z) =   \int\limits_0^{\infty}  dx_2\,dx_4\,dx_6 \,  e^{  - \sum\limits_{i=1}^3 (x_{2i}+ \frac{g}{2} \ln(x_{2i}/4)) - \frac{f}{2}(\ln(x_6) X+ \ln(x_2) Y+\ln(x_4) Z)}  \tilde q(x_4,x_5,x_6) \label{transform}  
\end{align}
where the measure (first factor in the exponent) is written to mimic the already existing measure over $x_1,x_3,x_5$ to make sure the full result is properly symmetric. Similarly, the factor $f/2$ in the second  factor in the exponential guarantees that the new $\ln(x_i) \ln(v_i)$ terms containing the even cross-ratios come with the same overall prefactor as their odd cousins in~(\ref{G6v3}). Note also that $\ln(x_6)$ will multiply $X$ which contains its two neighbors $\ln(x_{1=6+1})$ and $\ln(x_{5=6-1})$ and similarly for all other arguments. So in total we will get a beautiful symmetric chain of interactions and overall the only symmetry breaking term is $\tilde q(x_4,x_5,x_6)$! We should thus set it to a constant. Integrating (\ref{transform}) with $\tilde q$ equal to a constant indeed leads to the anticipated simple result~(\ref{finalresultppp}). This concludes our derivation. 
      \end{mdframed}

Putting everything together we thus find the final result for the full correlator in the light-like limit (and for general hexagon kinematics) as   
\begin{align}
\widetilde{G}_6 = &\underbrace{{\color{white}\!\!\!\! \Big(}\mathbb{W}(U_1,U_2,U_3)}_{\texttt{Renormalized Wilson loop}} \times \underbrace{ \exp\big( \sum\limits_{i} \frac{f}{16}\ln v_i \ln v_{i+3} -\frac{f}{8}\ln v_i \ln v_{i+1} + {{\frac{g- f \gamma_E}{4} \ln v_i}} \big) }_{\texttt{Sudokov Factor}} \times \label{G6v4}  \\
&\times\underbrace{ \mathcal{N} \left( \int\limits_0^{\infty}  \prod_{j=1}^6 dx_j   e^{  - \sum\limits_{i=1}^6 (x_{i}+ \frac{g}{2} \mathbf{ln}(x_{i})- \frac{f}{4}  \mathbf{ln}( x_i)  \mathbf{ln} (x_{i+1}) - \frac{f}{4}  \mathbf{ln}(x_{i})  \mathbf{ln}(v_{i}))+ \sum
\limits_{i} \frac{f \gamma_E}{4}  \mathbf{ln}(v_{i})} \right)}_{\texttt{Recoil }J}, \nn 
\end{align}
where $ \mathbf{ln}(x) = \ln (x) + \gamma_E$.

To obtain the full map between spinning three point functions and the Wilson loop we simply need to convert $\hat P$ to $\hat C$ using (\ref{PC}). In other words, we divide the result whence obtained by three large spin structure constants for a single spinning operator which were computed in \cite{AB}. This ratio nicely removes some of the gamma functions in (\ref{finalresultppp}) leading to our final main result 

      \begin{mdframed}[frametitle={Structure Constant/Wilson Loop duality},frametitlealignment=\centering,backgroundcolor=blue!6, leftmargin=0cm, rightmargin=0cm, topline=false,
	bottomline=false, leftline=false, rightline=false] 
\begin{equation}
\hat{C}^{\bullet \bullet \bullet}(J_i,\ell_i) 
=\underbrace{\mathcal{N} \prod_{i=1}^3 \Big( \frac{J_i \ell_i}{2\ell_{i+1} \ell_{i-1}}\Big)^{\frac{\gamma_i}{2}}}_{\texttt{conversion factor}}  \times   \mathbb{W}(U_1,U_2,U_3) \label{finalresult}.
\end{equation}
with the map between variables on both sides of this equation given by (\ref{lUmap}) or (\ref{Ulmap}). 
\end{mdframed}

\section{One-loop check and some speculative musings}
\label{oneloop}

\begin{figure}[t]
\centering
\includegraphics[width=\textwidth]{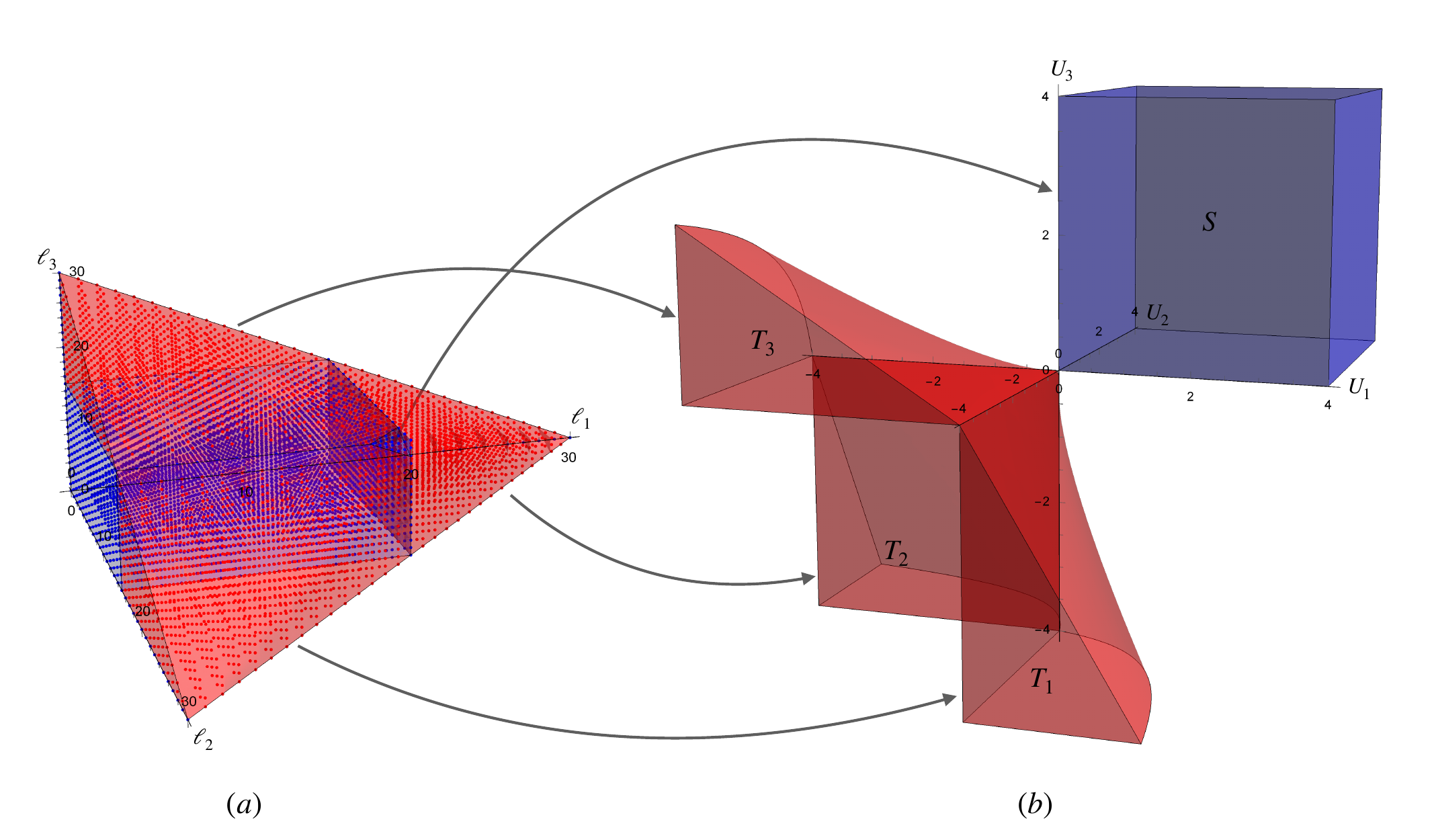}
\caption{The OPE data $\hat{C}^{\bullet\bullet\bullet}(J_1,J_2,J_3,\ell_1,\ell_2,\ell_3)$ can be plotted in the cross-ratio space $(U_1,U_2,U_3)$ if we map the $\ell_i$ and $J_i$ variables to the $U_i$ using (\ref{Ulmap}). The one loop structure constants have a good $J_i\to \infty$ limit in the blue region perfectly matching with the Wilson loop in the Euclidean space-like $(S)$ sheet. In contrast, the same structure constants blow up as $J_i \to \infty$ in the red region which would naively correspond to the Wilson loop in some Lorentzian time-like regions $(T_i)$. To reach these regions we should instead start in the blue region and take the large spin limit so that an emergent analytic structure arises with new branch cuts. We can then cross them by analytically continuing away from the blue region and in this way obtain a match beyond the Euclidean regime. The structure constant/Wilson loop duality is a cute concrete example where expansions and analytic continuations do not commute.  
}
\label{fig:LorentzVsEuclid}
\end{figure}

The structure constant variables $(J_1,J_2,J_3,\ell_1,\ell_2,\ell_3)$ are mapped into the Wilson loop cross-ratios $(U_1,U_2,U_3)$ through the map (\ref{Ulmap}). The $J_i$ are even non-negative integers and the $\ell_i$ are non-negative integers bounded by the condition that $\ell_i+\ell_j \le J_k$ with $i,j,k$ all different. For $J_1=J_2=J_3=30$ for instance we would have $7816$ discrete $\ell_j$ choices, each with its own structure constant. The map (\ref{Ulmap}) maps each one of these $\ell_j$ choices to a point in the cross-ratio space as depicted in figure \ref{fig:LorentzVsEuclid}.

The set of $\ell_k < J_i J_j/(J_i+J_j)$ covers the full space of positive real cross-ratios $U_j$ as represented in the figure \ref{fig:LorentzVsEuclid} by the blue dots/region. The remaining $\ell_k$'s cover three disjoint regions in cross-ratio space where one cross-ratio is positive and two are negative. (In the large spin limit of course.) The region of all positive cross-ratios can be called the \textit{space-like} region since it can be realized with all squared distances positive. The other three regions need some squared distances to be negative to get negative cross-ratios so we call them \textit{time-like regions}. (A beautiful detailed analysis of the geometry of the $U_i$ space for hexagonal Wilson loops is given in \cite{AGM}.)

We propose that as we take the large $J_k,\ell_k$ limit the structure constants in the space-like region ($S$) will nicely match -- according to (\ref{finalresult}) -- with the Wilson loop in the space-like (or Euclidean) sheet, where we start with all cusps space-like separated and do not cross any light-cone. Let us discuss a non-trivial one loop check of this proposal.

In perturbation theory we have $\hat{C}^{\bullet \bullet \bullet}(J_i,\ell_i) =1+\lambda c+\dots$, $\texttt{conversion factor}  = 1+ \lambda h+\dots $ and $\mathbb{W}(U_i) =1+\lambda w+\dots$ so that at one loop our prediction (\ref{finalresult}) simply translates into (up to an overall shift by a constant)
\beq
c(J_i,\ell_i)-h(J_i,\ell_i)=w\big(U_i(J_i,\ell_i)\big) \,. \la{proposal1Loop}
\eeq
The one loop Wilson loop is universal in any non-abelian gauge theory in the planar limit since it is given by a single gluon exchange from an edge of the hexagon to another. It reads \cite{AGM,bds,hexagonOPEpaper,origin}
\beq
w\big(U_1,U_2,U_3\big)= -4\pi^2 + 2\sum_{i=1}^3\text{Li}_2\left(1-1/U_i\right) \la{WLoneLoop}.
\eeq
This object -- in the space-like region where all $U_i$ are positive -- should emerge from the one loop structure constant of large spin operators. These are extracted from the OPE of the one loop correlation functions of six $20'$ operators in planar $\mathcal{N}=4$ SYM, see appendix \ref{appendix1Loopdata}. 
\vspace{0.5cm}

\begin{figure}[t]
\includegraphics[width=\textwidth]{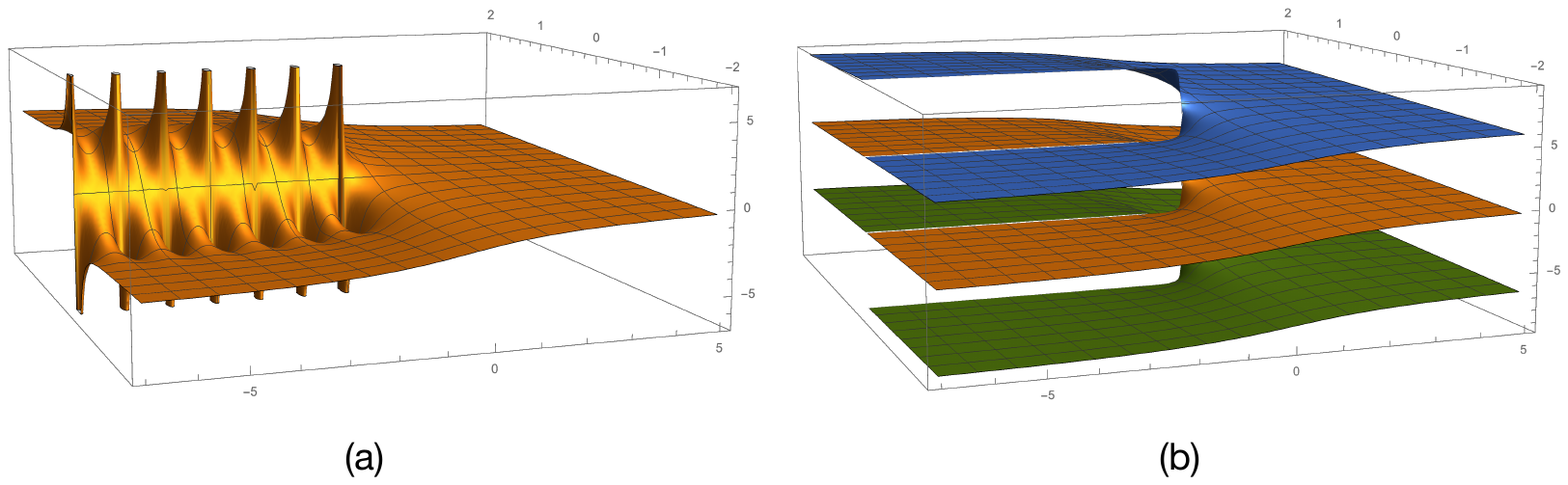}
\vspace{-7.7cm} \caption{\small \textbf{a)} (Imaginary part of) $H(z)$ where $H(z)$ are Harmonic numbers evaluate to rational numbers for $z$ positive integer and has poles at negative integers. For large argument it behaves as $\log(z)$. In other words, seen from far away the poles condense into a cut.  \textbf{b)} (Imaginary part of) $\log(z)$ lives in an infinite sheeted surface. The first one agrees with that of $H(z)$ for large arguments while the other sheets are \textit{emergent}. 
}
\label{SimplestHarmonicNumberExample}
\end{figure}

         \begin{mdframed}[frametitle={A speculative detour},frametitlealignment=\centering,backgroundcolor=blue!6, leftmargin=0cm, rightmargin=0cm, topline=false,
	bottomline=false, leftline=false, rightline=false] 

Before discussing the quantitative match of the structure constant and the Wilson loop we will open a speculative parentheses here. It can be skipped by the more orthodox readers. 

Note that the analytic structure of the structure constant before taking the large spin and large polarizations limit is strikingly different to that of the Wilson loop. 

The Wilson loop has a rich cut structure. In the physical sheet there are cuts at $U_i=0$ which need to be crossed to go from space-like to time-like configurations. These are the only singularities of the Wilson loop in the physical space-like sheet~\cite{hexagonOPEpaper}. If we cross the $U_i=0$ cuts we go to other sheets and do see other singularities most notably at $U_i=1$ but also at various other interesting locations, see e.g. \cite{Letters}.

 Instead, the structure constant are meromorphic functions of $\ell_i$ and $J_j$ with no cuts whatsoever -- see appendix \ref{appe} for explicit expressions full of Harmonic numbers, rational binomial sums and other similar meromorphic building blocks. They have poles at unphysical values of polarizations and spins. In the large $\ell_i$ and $J_i$ limit these poles condense; seen from far away they become cuts as illustrated in figure \ref{SimplestHarmonicNumberExample}. (This phenomenon of poles condensing into cuts is all over, most notably in Matrix model studies.) In other words, at finite $\ell_i, J_i$ there are no other sheets and no monodromies to be picked, only the space-like sheet exists. All other Lorentzian sheets are \textit{emergent}. They only appear in the \textit{semi-classical} limit of large $\ell_i,J_i$. As such, what we expect is that if we stay in the Euclidean regime $\ell_k < J_i J_j/(J_i+J_j)$ corresponding to the blue region in figure \ref{fig:LorentzVsEuclid} we should obtain a match with the Wilson loop in the large spin limit. But if we want to access other regions in the Wilson loop, the order of limits is key: We \textit{first} need to take the large spin and large polarization limit so cuts emerge; \textit{then} we analytically continue our structure constant through those cuts.

When doing a numerical comparison of the Wilson loop and the structure constants we observe an interesting phenomenon which seems to back this up. In the space-like $\ell_k < J_i J_j/(J_i+J_j)$ region the one loop structure constants $c$ are $O(1)$ numbers; as we increase the spin we observe that these numbers do approach the expected Wilson loop expression (\ref{WLoneLoop}). On the other hand, for $\ell_k > J_i J_j/(J_i+J_j)$ the structure constants $c$ become exponentially large real numbers which blow up as $J_j \to \infty$! This is in perfect synthony with the picture of the previous paragraph: to cross the cuts and reach the Lorentzian domain -encountering a complex valued finite Wilson loop - we must first go to a safe region in the physical sheet and then take a classical limit so the cuts appear in the first place. If we go to the cuts \textit{directly} in the structure constant side we encounter instead a divergence -- we could call it a firewall in analogy with black holes. In this black hole analogy, the smooth cuts with emerge in the classical limit resemble the smooth black hole horizons while the poles in the structure constants which we would only see through very sensitive experiments would be the analogue of the quantum black hole micro-states inner structure; some kind of fuzzball. 

A analytical toy model for this phenomenon is $\hat{C}^{\bullet \bullet \circ}$ in the large spin limit, discussed in detail in appendix \ref{twospinning}. In equation (\ref{beforeSP}) we obtain nice $O(1)$ expressions valid for $\ell < J_1 J_2/(J_1+J_2)$ with emergent branch points at $\ell = J_1 J_2/(J_1+J_2)$ which should be thought of as analogues for the $U=0$ light-cone singularities of the hexagonal Wilson loop. On the other hand, for $\ell >  J_1 J_2/(J_1+J_2)$ the one-loop corrections become exponentially divergent, see equation (\ref{afterSP}). In fact, extending the black hole analogy, one must be careful when using these limits to compute observables that probe the ``horizon" or ``interior" regions. For example, the equal spins sum over $\ell$ of the three point function is finite in the large spin limit -- given by (\ref{sum2spin}) in the appendix -- but it is not purely captured by the naive large $\ell$, $J$ limit with $\ell/J$ fixed. Indeed, if one first takes the large spin limit, a non integrable expression is obtained in the ``interior" region. This singularity can be thought of as a ``UV" divergence that is regulated by finer corrections coming from the microscopical structure of three point functions.

{\centering{ \textbf{End of speculative detour. }} \\}
     \end{mdframed}

We could not completely fix the analytic form of the one loop structure constants $c(J_i,\ell_i)$ and so we could not establish (\ref{proposal1Loop}) fully.  
Instead we expanded the Wilson loop around the so-called origin limit \cite{origin} corresponding to small cross-ratios. Note that we have
\beq
w\big(U_1,U_2,U_3\big)=  \sum_{i=1}^3 \ln^2\left(U_i\right) + \sum_{i=1}^3 \ln\left(U_i\right) \mathcal{A}_i + \mathcal{B}  \la{repAB}
\eeq
where $\mathcal{A}_i$ and $\mathcal{B}$ have regular Taylor expansions around the origin $U_1=U_2=U_3=0$.\footnote{Explicitly, $\mathcal{A}_i=2 \ln(1-U_i)$ and $\mathcal{B}=2 \text{Li}_2(U_1)+2 \text{Li}_2(U_2)+2 \text{Li}_2(U_3)-5\pi^2$.} For instance 
\beq
\mathcal{B}=-5 \pi ^2+\frac{2}{1} \sum_{i} U_i+\frac{2}{4} \sum_{i} U_i^2+\frac{2}{9}
 \sum_{i} U_i^3+\frac{2}{16}  \sum_{i} U_i^4+\dots  \label{Bexp}
\eeq

The representation (\ref{repAB}) makes manifest the branch-cuts at $U_i=0$ of the Wilson loop. In the structure constant side, to make contact with the Wilson loop as an expansion around the origin we should consider the limit of very large spin and polarizations but very small ratios of the two,
\beq
\ell_i \gg 1 \,, \qquad J_i \gg 1 \,, \qquad \ell_i/ J_j \ll 1\,,
\label{defOriginLimit}
\eeq
indeed, in this regime we easily see that the cross-ratios obtained through (\ref{Ulmap}) are very small, for example:
\beq
U_1=\frac{\ell_1\ell_3}{J_2^2}+\frac{\ell_1^2\ell_3}{J_2^3}+\frac{\ell_1^2\ell_3}{J_2^2J_3}+\frac{\ell_1\ell_3^2}{J_2^3}+\frac{\ell_1\ell_3^2}{J_1J_2^2}+\dots \,, \;\ln(U_1)=\ln\left(\frac{\ell_1\ell_3}{J_2^2}\right)+\frac{\ell_1}{J_2}+\frac{\ell_1}{J_3}+\frac{\ell_3}{J_1}+\frac{\ell_3}{J_2}+\dots\,.
\label{smallCR}
\eeq

When matching the one-loop correlation function $c$ with the Wilson loop $w$ the various logs arising in the large spin limit of the structure constants should match the explicit logs in (\ref{repAB}) while the powerlaw corrections in $\ell_j/J_k$ should be matched with the series expansion of $\mathcal{B}$ and $\mathcal{A}_i$ for small cross-ratios.\footnote{In the structure constant there are also terms like $\ell_1/J_2^3$ and so on which have less powers of $\ell$'s in the numerator compared to powers of $J$'s in the denominator; we call these terms \textit{unbalanced}. The unbalanced terms vanish in the large spin/large polarization limit so we are insensitive to them when testing the WL/Correlation function duality. In other words, the structure constants contain way more information than the Wilson loop. We can think of them as an off-shell quantum version of the Wilson loop which reduced to the Wilson loop in a classical limit where we keep balanced terms only such as the ones in the expansions~(\ref{smallCR}).} If we can match all terms in these Taylor expansions we would establish (\ref{proposal1Loop}) completely. We almost did it. We matched all terms in the expansion of $\mathcal{A}_i$ (see discussion around (\ref{T1Terms}) in the appendix \ref{appe}) and we matched the first 873 terms in the expansion of $\mathcal{B}$ once we translate (\ref{Bexp}) into small ratios expansions as (\ref{smallCR}) to more easily compare with the structure constants (see discussion around (\ref{T2Terms}) in the appendix \ref{appe}). This is more than plenty to leave zero doubt in our mind that (\ref{proposal1Loop}) holds. To fully establish it we would need to finish the full analytic determination of the structure constants which translate into finding a closed expression to the very simple remaining $\beta$ constants discussed around (\ref{TabBetas}) in appendix \ref{appe}. It would be very nice to find these constants. One reason is to conclude this analytic comparison but a perhaps even more interesting reason would be to analytically understand all the various speculative remarks about the behavior of the structure constants inside and outside the Euclidean regime which we mused about in the speculative detour above.

%

\begin{figure}[t]
\includegraphics[width=\textwidth]{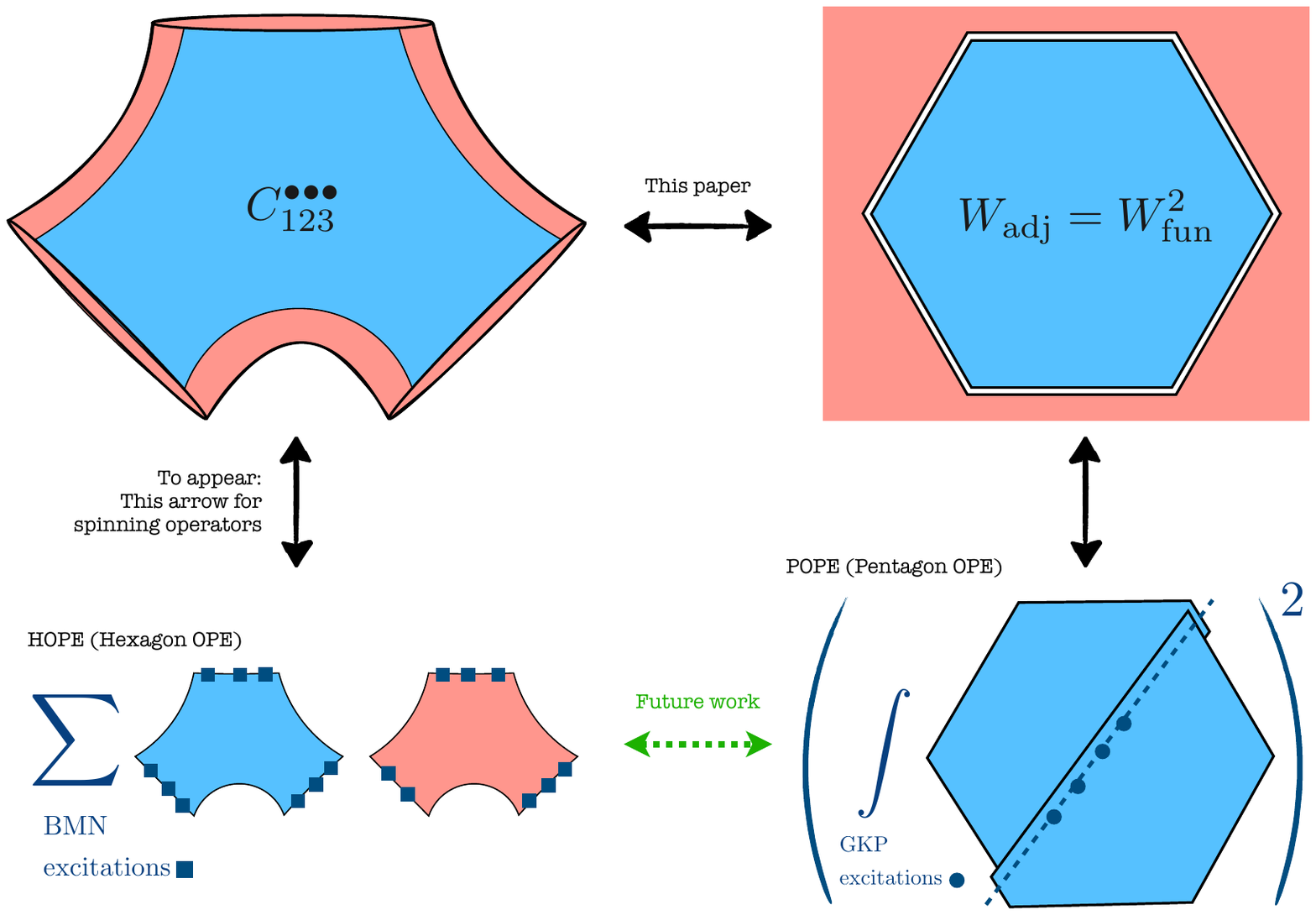}
\vspace{-1.7cm} \caption{\small \textbf{Top arrow:} Large spin three-point function/hexagon Wilson loop duality \cite{nullOPE}. \textbf{Left arrow:} Three point functions can be decomposed in terms of two hexagons \cite{hexagons}. For spinning operators the necessary formalism is cleaned up in \cite{Paper2}. \textbf{Right arrow:} Wilson loops can be decomposed in terms of two pentagons \cite{POPE}. \textbf{Bottom arrow: }The top duality hints at a transmutation of hexagons into pentagons in the large spin limit. Would be fascinating to find out how this works. It might lead to a unified integrability description of open and closed strings in AdS/CFT.
}
\label{summaryFig}
\end{figure}

\section{Discussion}

This paper concerns the duality relation depicted at the top of figure \ref{summaryFig}. On the top left corner we have three point functions of three twist-two operators with large spins~$J_i$ and with polarizations tensor structures parametrized by $\ell_i$. On the top right corner we have a renormalized Wilson loop parametrized by three finite conformal cross-ratios $U_i$. Our main result is (\ref{finalresult}) which precisely links these two quantities with a precise kinematical dictionary.\footnote{Key in deriving this result was the so-called snowflake decomposition of six point correlation function. It is an interesting open problem to use instead the comb decomposition of a six point correlation function and arrive at the Wilson loop limit. We hope to report on this problem in the future.

The method used in this paper can also applied to derive a link between three point functions of two spinning operators and the expectaction value of a square Wilson loop and a local operator \cite{Berenstein:1998ij,Alday:2011ga}.}

Armed with a precise dictionary for the kinematics we can now attack the dynamics of this problem from an integrability perspective. 



Three point functions of three excited two-two operators  (each parametrized by $J_i$ integrability magnon excitations) can be decomposed in terms of two hexagons \cite{hexagons}. When cutting each operator into two these excitations can end up on either hexagon; we must sum over where they end up as indicated in the bottom left corner of figure \ref{summaryFig}.\footnote{In principle we should also integrate over all possible mirror states at the three lines where the two hexagons are glued to each other. We are ignoring this extra contribution. We believe it is subleading at large spin when the effective size of all operators is very large. We are currently trying to check this by an explicit finite size computation.}  The larger the spin, the more excitations we have and thus the scarier are these sums. In the large spin limit they ought to simplify and give rise to a Wilson loop (an adjoint Wilson loop or the square of a fundamental one). In turn, the Wilson loop can be obtained by gluing together two pentagons and summing over all possible virtual particles therein \cite{POPE}. So the sum over hexagon's with their large number of BMN physical excitations should somehow transmute into a sum over pentagons with a sum over GKP virtual excitations. To attack this fascinating alchemy exercise, we need to understand how to polarize the hexagon OPE expansion for spinning operators (all examples so far were for scalar structure constants or spinning structure constants with a single tensor structure). This is the subject of our upcoming work~\cite{Paper2}.\footnote{Related to that, in appendix \ref{appendix1Loop} we extract data for $C^{\bullet \bullet \bullet}_{123}$ at one loop in $\mathcal{N}=4$ SYM generalizing previous results by Marco Bianchi in \cite{Marco}; as usual, this data will be very useful in testing any integrability based approaches.
}


\section*{Acknowledgements}
We would like to thank Benjamin Basso, Marco Bianchi, Frank Coronado for illuminating discussions. 
Research at the Perimeter Institute is supported in part by the Government of Canada 
through NSERC and by the Province of Ontario through MRI. This work was additionally 
supported by a grant from the Simons Foundation (Simons Collaboration on the Nonperturbative Bootstrap \#488661 and \#488637) and ICTP-SAIFR FAPESP grant 2016/01343-7 and FAPESP grant 2017/03303-1. The work of P.V. was partially supported by the Sao Paulo Research Foundation (FAPESP) under Grant No. 2019/24277-8.  Centro de Fisica do Porto is partially funded by Funda\c{c}ao para a Ciencia e a Tecnologia (FCT) under the grant UID04650-FCUP. The work of C.B. was supported by the Sao Paulo Research Foundation (FAPESP) under Grant No. 2018/25180-5.

\newpage
\appendix 

\section{Spinors}
\label{app:spinnors}

Equation (\ref{3pt}) provides a manifestly covariant expression for the three point function. However, with the prospects of fitting this work in the context of $\mathcal{N}=4$ integrability, it is useful to express the three point function in a convenient conformal frame following \cite{bootstrapequations4d}. In $-+++$ signature we choose
\begin{align}
x_1 = (0,0,0,0)\,, \qquad  x_2 = (0,0,1,0) \,, \qquad 
x_3 = (0,0,\mathcal{L},0)
\end{align}
and consider the rescaled correlator
\begin{equation}
\mathbb{C}(\epsilon_1, \epsilon_2, \epsilon_3) \equiv \lim_{\mathcal{L} \rightarrow \infty} \mathcal{L}^{2\Delta_3}\langle \mathcal{O}_1 (x_1, \epsilon_1), \mathcal{O}_2 (x_2, \epsilon_2) , \mathcal{O}_3 (x_3, \epsilon_3) \rangle.
\end{equation}
Besides choosing a frame, in the context of integrability, it is useful to express the correlator in terms of polarization spinors. That is, to each operator $\mathcal{O}_i$ we assign auxiliary spinors ${L_i}^\alpha$ and ${R_i}_{\dot{\beta}}$. These are related to the polarization vectors by
\begin{equation}
\epsilon_i^\mu = {R_i}_{\dot{\beta}}  \bar{\sigma}^{\mu \dot{\beta} \alpha} {L_i}_\alpha ,
\end{equation}
where in our conventions the sigma matrices $ \sigma^{\mu}_{{\alpha} \dot{\alpha}}$,   $\bar{\sigma}^{\mu \dot{\alpha} \alpha}$  are given by
\begin{equation}
\sigma^0 = 
\begin{pmatrix}
-1 & 0 \\
0 & -1 
\end{pmatrix}, \qquad 
\sigma^1 = 
\begin{pmatrix}
0 & 1 \\
1 & 0 
\end{pmatrix}, \qquad 
\sigma^2 = 
\begin{pmatrix}
0 & -i \\
i & 0 
\end{pmatrix}, \qquad 
\sigma^3 = 
\begin{pmatrix}
1 & 0 \\
0 & -1 
\end{pmatrix}, \qquad 
\end{equation}
$\bar{\sigma} = (\sigma_0, -\sigma_1,-\sigma_2,-\sigma_3)$, and indices are raised and lowered with
\begin{equation}
-\epsilon_{\alpha \beta} = \epsilon^{\alpha \beta} = -\epsilon_{\dot{\alpha} \dot{ \beta}} =  \epsilon^{\dot{\alpha} \dot{ \beta}} =\begin{pmatrix}
0 & 1 \\
-1 & 0 
\end{pmatrix}.
\end{equation}
One of the very nice features of this conformal frame is how clean the invariant structures $H$ and $V$ in (\ref{3pt}) become. In terms of the left and right spinors they simply read
\begin{equation}
H_{ij} = \langle L_i , R_j \rangle \langle L_j , R_i \rangle, \qquad V_{i,jk} = \langle L_i, R_i\rangle,\qquad \langle L_i, R_j \rangle \equiv i {R_j}_{\dot{\alpha}} \bar{\sigma}_{2}^{\dot{\alpha} \alpha}   {L_i}_\alpha    \,.
\end{equation}

A general spinning three point function can then be cast as linear combination of monomials made out of these brackets. For instance, translating the results of appendix \ref{appendix1Loop} for the case of three spinning operators with spin $2,4,6$ respectively we can write 
\begingroup\makeatletter\def\f@size{6}\check@mathfonts
\def\maketag@@@#1{\hbox{\m@th\large\normalfont#1}}%
\begin{align}
    C^{\bullet\bullet\bullet}_{246} &= \left(\frac{1}{84\sqrt{55}}\right)\Bigg(\langle 1 1\rangle \langle 1 
  3\rangle\langle 2 
  2 \rangle^2 \langle 2 
  3\rangle^2 \langle 3 
  1\rangle \langle 3 
  2\rangle^2 \langle 3 
  3\rangle^3\left( 720-\frac{2480321}{1155}g^2\right)+\langle 1 1\rangle  \langle 1 
  2\rangle  \langle 2 
  1\rangle  \langle 2 
  2\rangle ^3 \langle 3 
  3\rangle ^6\left(8-\frac{1202701}{17325}g^2\right)+\nonumber \\
  &+\langle 1 1\rangle  \langle 1 
  3\rangle  \langle 2 
  2\rangle ^3 \langle 2 
  3\rangle  \langle 3 
  1\rangle  \langle 3 
  2\rangle  \langle 3 
  3\rangle ^4 \left(240-\frac{600189}{385}g^2\right)
  +\langle 1 
  1\rangle ^2 \langle 2 
  2\rangle ^2 \langle 2 
  3\rangle ^2 \langle 3 
  2\rangle ^2\langle 3 
  3\rangle ^4 \left( 90-\frac{822427}{1155}g^2\right)+ \nonumber \\
  &+\langle 1 2\rangle  \langle 1 
  3\rangle  \langle 2 
  1\rangle  \langle 2 
  2\rangle  \langle 2 
  3\rangle ^2 \langle 3 
  1\rangle  \langle 3 
  2\rangle ^2 \langle 3 
  3\rangle ^3 \left(1440-\frac{811882}{1155}g^2\right)+\langle 1 1\rangle  \langle 1 
  2\rangle  \langle 2 
  1\rangle  \langle 2 
  2\rangle ^2 \langle 2 
  3\rangle  \langle 3 
  2\rangle  \langle 3 
  3\rangle ^5\left(144-\frac{1335487}{1925}g^2\right)+\nonumber \\
  &+\langle 1 2\rangle  \langle 1 
  3\rangle  \langle 2 
  1\rangle  \langle 2 
  2\rangle ^2 \langle 2 
  3\rangle  \langle 3 
  1\rangle  \langle 3 
  2\rangle  \langle 3 
  3\rangle ^4\left(720-\frac{179332}{385}g^2\right)+\langle 1 1\rangle \langle 1 
  2\rangle \langle 2 
  1\rangle \langle 2 
  2\rangle \langle 2 
  3\rangle^2 \langle 3 
  2\rangle^2 \langle 3 
  3\rangle^4\left(360-\frac{433393}{1155}g^2\right)+ \nonumber \\
  &+\langle 1 1\rangle ^2 \langle 2
   2\rangle ^3 \langle 2 
  3\rangle  \langle 3 
  2\rangle  \langle 3 
  3\rangle ^5\left(24-\frac{694912}{1925}g^2\right)+\langle 1 1\rangle  \langle 1 
  3\rangle  \langle 2 
  2\rangle ^4 \langle 3 
  1\rangle  \langle 3 
  3\rangle ^5\left(12-\frac{716273}{5775}g^2\right)+ \nonumber  \\
  &+\langle 1 1\rangle ^2 \langle 2
   2\rangle ^4 \langle 3 
  3\rangle ^6\left(1-\frac{814939}{34650}g^2\right)+\langle 1 2\rangle  \langle 1 
  3\rangle  \langle 2 
  1\rangle  \langle 2 
  2\rangle ^3 \langle 3 
  1\rangle  \langle 3 
  3\rangle ^5\left(48 +\frac{36268}{5775}g^2\right) +\nonumber \\
  &+\langle 1 3\rangle ^2 \langle 2
   3\rangle ^4 \langle 3 
  1\rangle ^2 \langle 3 
  2\rangle ^4\left(15 +\frac{4261}{231}g^2\right)+\langle 1 
  2\rangle ^2\langle 2 
  1\rangle ^2\langle 2 
  2\rangle ^2\langle 3 
  3\rangle ^6\left(6 +\frac{219941}{5775}g^2\right) +\nonumber \\
  &+\langle 1 3\rangle ^2 \langle 2
   2\rangle ^4 \langle 3 
  1\rangle ^2 \langle 3 
  3\rangle ^4\left(15 +\frac{25231}{462}g^2\right)+\langle 1 1\rangle ^2 \langle 2
   2\rangle  \langle 2 
  3\rangle ^3 \langle 3 
  2\rangle ^3 \langle 3 
  3\rangle ^3\left(80 +\frac{670748}{3465}g^2\right) +\nonumber \\
  &+\langle 1 2\rangle ^2 \langle 2
   1\rangle ^2 \langle 2 
  2\rangle  \langle 2 
  3\rangle  \langle 3 
  2\rangle  \langle 3 
  3\rangle ^5\left(72 +\frac{572534}{1925}g^2\right)+\langle 1 
  2\rangle ^2 \langle 2 
  1\rangle ^2\langle 2 
  3\rangle ^2\langle 3 
  2\rangle ^2\langle 3 
  3\rangle ^4\left(90 +\frac{422663}{1155}g^2\right) +\nonumber \\
  &+\langle 1 2\rangle  \langle 1 
  3\rangle  \langle 2 
  1\rangle  \langle 2 
  3\rangle ^3 \langle 3 
  1\rangle  \langle 3 
  2\rangle ^3 \langle 3 
  3\rangle ^2\left(480 +\frac{433666}{1155}g^2\right)+\langle 1 
  3\rangle ^2 \langle 2 
  2\rangle  \langle 2 
  3\rangle ^3\langle 3 
  1\rangle ^2\langle 3 
  2\rangle ^3\langle 3 
  3\rangle \left(240 +\frac{97282}{231}g^2\right) +\nonumber \\
  &+\langle 1 
  3\rangle ^2 \langle 2 
  2\rangle ^3 \langle 2 
  3\rangle  \langle 3 
  1\rangle ^2\langle 3 
  2\rangle \langle 3 
  3\rangle ^3\left(240 +\frac{33146}{77}g^2\right)+\langle 1 1\rangle ^2 \langle 2
   3\rangle ^4 \langle 3 
  2\rangle ^4 \langle 3 
  3\rangle ^2\left(15 +\frac{520958}{1155}g^2\right) +\nonumber \\
  &+\langle 1 3\rangle ^2 \langle 2
   2\rangle ^2 \langle 2 
  3\rangle ^2 \langle 3 
  1\rangle ^2 \langle 3 
  2\rangle ^2 \langle 3 
  3\rangle ^2\left(540 +\frac{53288}{77}g^2\right)+\langle 1 1\rangle  \langle 1 
  2\rangle  \langle 2 
  1\rangle \langle 2 
  3\rangle ^3\langle 3 
  2\rangle ^3\langle 3 
  3\rangle ^3\left(160 +\frac{2899591}{3465}g^2\right) +\nonumber \\
  &+\langle 1 1\rangle  \langle 1 
  3\rangle  \langle 2 
  3\rangle ^4 \langle 3 
  1\rangle  \langle 3 
  2\rangle ^4 \langle 3 
  3\rangle \left(60 +\frac{984272}{1155}g^2\right)+\langle 1 1\rangle  \langle 1 
  3\rangle  \langle 2 
  2\rangle  \langle 2 
  3\rangle ^3 \langle 3 
  1\rangle  \langle 3 
  2\rangle ^3 \langle 3 
  3\rangle ^2\left(480 +\frac{1301071}{1155}g^2\right)\Bigg)
\end{align}\endgroup
where $\langle ij \rangle=\langle L_i,R_j \rangle$.

Note that by simply looking at the various homogeneous degrees in $L_i$ and $R_i$ we can automatically infer the three spins of this correlator. A very concrete test of the spinning hexagonalization cleaned up in \cite{Paper2} is to reproduce this equation from the hexagon formalism. 

\section{The $\ell(\epsilon)$ map} 

 In the limit of large spin the structure constants ${C}^{\bullet  \bullet \bullet}$ are exponentially small.  However, as remarked in section \ref{oneloop},  $\hat{C}^{\bullet   \bullet \bullet}$ suffers from a Stokes-like phenomenon: in the space-like region it is of order one, while in the time-like region it diverges exponentially. These leads to a dramatic simplification in the sum of (\ref{3pt}) in the large spin limit. Provided the tree-level decay combined with the chemical potentials $T_{i,jk} \equiv \frac{H_{jk}}{V_{j,k i} V_{k,i j}}$ are enough to  suppress the contribution from the time-like regions, (\ref{3pt}) reduces to a saddle point computation governed by tree-level\footnote{In the toy model case of two spinning operators at one-loop discussed in the previous section, the sum over tensor structures also simplifies to a saddle computation under the same conditions. There, the chemical potential can suppress the exponential divergences of the one-loop structure constants provided $T_{3,1 2}<1$.}. The loop corrections are then simply evaluated \textit{at} the saddle location, in which they are of order one.\footnote{We verified this statement numerically in perturbation theory.}
 
In this regime, we thus conclude that 
\beq
\frac{\langle \mathcal{O}_{J_1} (x_1, \epsilon_1) \mathcal{O}_{J_2} (x_2, \epsilon_2)  \mathcal{O}_{J_3} (x_3, \epsilon_3) \rangle}{\langle \mathcal{O}_{J_1} (x_1, \epsilon_1) \mathcal{O}_{J_2} (x_2, \epsilon_2)  \mathcal{O}_{J_3} (x_3, \epsilon_3) \rangle_\text{tree level}} = \hat C^{\bullet  \bullet \bullet} \qquad \text{evaluated at $\ell_i$ given by} \qquad  \frac{\partial W}{\partial{\ell_i}} = 0 \nn
\eeq
where
\begin{equation}
e^{W} \equiv  \left( C^{\bullet \bullet \bullet}_\text{tree level} \simeq \prod_i \frac{   \big(J_i+\ell _i-\sum\limits_k \ell _k\big){}^{\sum\limits_k \ell _k-J_i-\ell _i-\frac{1}{2}}}{\pi^{\frac{3}{4}}  \ell _i^{2
   \ell _i+1}\,e^{\sum\limits_k \ell _k-3 \ell _i}\,2^{J_i+1}J_i^{-J_i-\frac{3}{4}}} \right){V_{1,23}^{J_1 - l_2 - l_3} }{V_{2,31}^{J_2 - \ell_3 - \ell_1}} {V_{3,12}^{J_3 - \ell_1 - \ell_2}} {H_{23}^{\ell_1}}  {H_{31}^{\ell_2}}  {H_{12}^{\ell_3}}  \,.\nn
\end{equation} 
The critical points of the potential $W$ in the large spin limit take a remarkably simple form 
\begin{align}
\frac{H_{2,3}}{V_{2,3 1} V_{3,1,2}} & = \frac{\ell_1^2}{(J_3-\ell_1-\ell_2)(J_2-\ell_1-\ell_3)} \nonumber \\
\frac{H_{3,1}}{V_{3,1 2} V_{1,2 3}} & = \frac{\ell_2^2}{(J_1-\ell_2-\ell_3)(J_3-\ell_2-\ell_1)} \label{saddlel} \\
\frac{H_{1,2}}{V_{1,2 3} V_{2,3 1}} & = \frac{\ell_3^2}{(J_2-\ell_3-\ell_1)(J_1-\ell_3-\ell_2)}\nonumber
\end{align}
where $J_i$ and $\ell_i$ are both large and of the same order. This constitutes the sought after $\epsilon(\ell)$ map.\footnote{Of course, we can not fix the polarization themselves but only meaningful conformal invariant combinations in the left hand side of relation (\ref{saddlel}). In practice we can however pick a particular conformal frame and use (\ref{saddlel}) to define a realization $\epsilon_i(\ell_1,\ell_2,\ell_3)$. This particular realization is what is most convenient when dealing with these structure constants from an integrability perspective as explained in \cite{Paper2}.} We emphasize this map should be understood to hold in the space-like region corresponding to the image covered by positive $U$'s through the $\ell(U)$ map (\ref{lUmap}). To access the time-like regions, one must analytically continue away from the well controlled Euclidean region.

Combining (\ref{saddlel}) with (\ref{lUmap}) we can also translate our map into spinor variables. We get
\begin{equation}
{\color{blue}{\frac{ \langle L_{i+1}, R_{i+1}\rangle \langle L_{i+2}, R_{i+2}\rangle}{ \langle L_{i+1} , R_{i+2} \rangle \langle L_{i+2} , R_{i+1} \rangle}}} = \left(R_{i+1; i+2 \hspace{2pt} i} - \frac{1+R_{i+1; i+2 \hspace{2pt} i}}{1+R_{i; i+2 \hspace{2pt} i+1}}\right) \left(R_{i+2; i+1 \hspace{2pt} i} - \frac{1+R_{i+2; i+1 \hspace{2pt} i}}{1+R_{i; i+1 \hspace{2pt} i+2}}\right) \la{epsilonMap}
\end{equation}
with $i=1,2,3$, all indices taken modulo $3$ and where \beq R_{a; b\hspace{2pt} c} = \frac{{\color{red}{\frac{J_b}{J_a}}}+{\color{red}{\frac{J_c}{J_a}}}\sqrt{{\color{ForestGreen}{\frac{\tilde{U}_c}{\tilde{U}_a\tilde{U}_b}}}}}{{\color{red}{\frac{J_c}{J_a}}}+{\color{red}{\frac{J_b}{J_a}}}\sqrt{{\color{ForestGreen}{\frac{\tilde{U}_b}{\tilde{U}_a\tilde{U}_c}}}}}, \qquad \tilde{U}_1 = U_2, \qquad \tilde{U}_2 = U_1, \qquad\tilde{U}_3 = U_3.\eeq 

Note that there are more variables in the three point function side of the duality so we have some freedom on how to approach the duality. Only ratios matter: We can take the large spin limit of the three point function with all spins (approximately) equal, for instance. Then all the (red) ratios of spins evaluate to~$1$ and this already simple relation simplifies even more.

\section{Cross-ratios}
\label{appCR}
\begin{figure}[t]
\includegraphics[width=\textwidth]{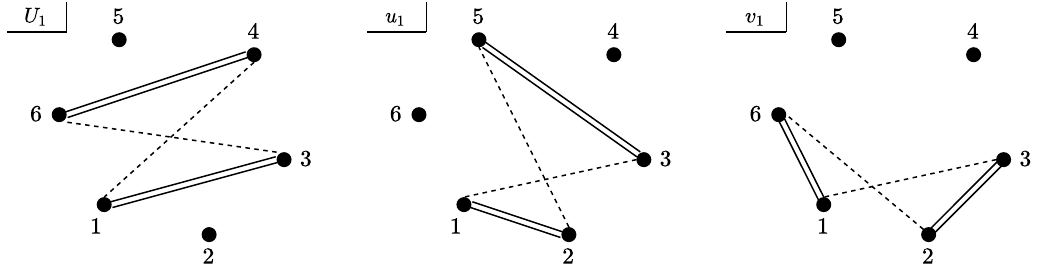}
\caption{\small The six point function has nine independent cross-ratios. We pick three of them to be $U_1$, represented here, plus its two cyclic images and the remaining six are either the $u_1$ plus its cyclic images or $v_1$ and its cyclic images, both represented above. We convert these pictures into the formulae in the text by writing the cross-ratio as the (product of the square of the) distances represented by the solid lines divided by the ones associated with the dashed lines.}
\label{CrossRatios}
\end{figure}

The cross-ratios used to write the six point correlation function are presented in figure \ref{CrossRatios}.
We have 
\begin{equation}
u_1 = \frac{{\color{blue} x_{12}^2} x_{35}^2}{x_{13}^2 x_{25}^2}\,, \quad u_2 = \frac{{\color{red} x_{23}^2} x_{46}^2}{x_{24}^2 x_{36}^2}\,, \quad u_3= \frac{{\color{blue} x_{34}^2} x_{51}^2}{x_{35}^2 x_{41}^2}\,, \quad 
u_4 = \frac{{\color{red} x_{45}^2 }x_{62}^2}{x_{46}^2 x_{52}^2}\,, \quad u_5 =  \frac{{ \color{blue} x_{56}^2}  x_{13}^2}{x_{51}^2 x_{63}^2}\,, \quad u_6 = \frac{{\color{red} x_{61}^2} x_{24}^2}{x_{62}^2 x_{14}^2}
\end{equation}
with $x_{ij}^2=(x_i-x_j)^2$. 
This set of cross-ratios $u_i$ has a single null distance in the numerator when we send $x_{i,i+1}^2\to 0$. This is quite convenient for taking one null limit at a time. 

For instance, if we take ${\color{blue} 12}$, ${ \color{blue}34}$ and $\color{blue} 56$ to become null that sets the odd cross-ratios $u_1, u_3, u_5 \to 0$. That is represented in figure \ref{snow}a. This is a so-called single light-like OPE. In this null limit we have $(x_1-x_2)^2\to 0$ since $x_1$ and $x_2$ are becoming null separated without necessarily colliding with each other. We could make this distance zero with the stronger condition $x_2\to x_1$ corresponding to the euclidean OPE. If we take all pairs $12$, $34$ and $56$ to collide in this Euclidean sense we set not only the odd cross-ratios to zero, $u_1, u_3, u_5 \to 0$, but we should expand the even ones around one, $u_2, u_4, u_6 \to 1$, as represented in figure \ref{snow}b. A very different limit we could take is to keep $12$, $34$ and $56$ null and then send the remaining consecutive pairs of points $\color{red} 23$, $\color{red} 45$ and $\color{red}  61$ to be null so that in total all consecutive points are null drawing a full closed polygon as represented in figure \ref{snow}c. In that case we set all the odd and even cross-ratios to zero $u_i \to 0$. This very Lorentzian limit is also called a double light-like OPE limit. 

In this paper we always take $u_1, u_3, u_5$ to zero first. This projects into leading twist operators in the corresponding three OPE channels as represented in figure \ref{snow}a. In the main text we then take the remaining $u_2,u_4,u_6$ to zero to construct a full null polygonal configuration. This projects further into large spin exchanges. In appendix \ref{appendix1Loopdata}, instead, we expand around the Euclidean limit where $u_2, u_4,u_6 \to 1$ to extract OPE data for finite spins. 

Another important set of cross-ratios  is
\begin{equation}
v_1 = \frac{ \color{red}  x_{61}^2x_{23}^2}{x_{62}^2 x_{13}^2}\,, \quad v_2 = \frac{\color{blue} x_{12}^2x_{34}^2}{x_{13}^2 x_{24}^2}\,, \quad v_3 = \frac{ \color{red}  x_{23}^2x_{45}^2}{x_{24}^2 x_{35}^2}\,, 
\quad v_4 = \frac{\color{blue} x_{34}^2x_{56}^2}{x_{35}^2 x_{46}^2}\,, \quad v_5 = \frac{\color{red} x_{45}^2x_{61}^2}{x_{46}^2 x_{51}^2}\,, \quad v_6 = \frac{\color{blue}  x_{56}^2x_{12}^2}{x_{51}^2 x_{62}^2}\,,
\end{equation}
which have two vanishing distances in the null limit. Because of these two distances, taking (either Euclidean or Lorentzian) OPE limits in a sequential way as discussed above is murkier in the $v_j$ language. They have, however, the big advantage of being very local and symmetric, more so that the $u_j$ as can be clearly seen in figure \ref{CrossRatios}. It is in these local variables that the important recoil effect introduced in \cite{Alday:2010zy} is expressed. We will thus use the $u_j$'s for most derivations but switch to the $v_j$'s when imposing the required symmetries of the final results to bootstrap the correlators.

The remaining three cross-ratios $(U_1,U_2,U_3)$ which parametrize the six point function are
\begin{equation}
U_1 = \frac{x_{13}^2 x_{46}^2}{x_{14}^2 x_{36}^2}\,, \qquad U_2 = \frac{x_{24}^2 x_{51}^2}{x_{25}^2 x_{41}^2}\,,  \qquad U_3 = \frac{x_{35}^2 x_{62}^2}{x_{36}^2 x_{52}^2}\,. 
\end{equation}
These cross-ratios remain finite in the double light-cone limit. They parametrize the resulting null hexagons. (In the triple Euclidean OPE we have $U_i \to 1$.)


\section{Conformal block integrand $\mathcal{F}$} \la{Fappendix}
The light cone six point snowflake conformal block governing the exchange of leading twist single trace operators and all its descendents admits a simple triple integral representation~\cite{Multi} which we quote here for completeness. The integrand $\mathcal{F}$ in (\ref{start}) reads
\begin{align}
\mathcal{F} & = \frac{\Gamma\left(2J_1+\gamma_1+2\right)}{\Gamma\left(J_1+\frac{\gamma_1}{2}+1\right)^2}\frac{\Gamma\left(2J_2+\gamma_2+2\right)}{\Gamma\left(J_2+\frac{\gamma_2}{2}+1\right)^2}\frac{\Gamma\left(2J_3+\gamma_3+2\right)}{\Gamma\left(J_3+\frac{\gamma_3}{2}+1\right)^2} \frac{\textcolor{red}{u_2}\textcolor{red}{u_4}\textcolor{red}{u_6}}{\textcolor{red}{u_1}\textcolor{red}{u_3}\textcolor{red}{u_5}}\, \textcolor{blue}{U_1} \textcolor{blue}{U_2} \textcolor{blue}{U_3} \,\textcolor{red}{u_1}^\frac{\gamma_1}{2} \textcolor{red}{u_3}^\frac{\gamma_2}{2} \textcolor{red}{u_5}^\frac{\gamma_3}{2} \nonumber \\
& \times (\textcolor{blue}{U_1}-\textcolor{red}{u_2})^{\ell_3}(\textcolor{blue}{U_2}-\textcolor{red}{u_6})^{\ell_2}(\textcolor{blue}{U_3}-\textcolor{red}{u_4})^{\ell_1}\textcolor{blue}{U_1}^{J_2+\frac{\gamma_1}{2}+\frac{\gamma_2}{2}-\frac{\gamma_3}{2}}\textcolor{blue}{U_2}^{J_1+\frac{\gamma_1}{2}-\frac{\gamma_2}{2}+\frac{\gamma_3}{2}}\textcolor{blue}{U_3}^{J_3-\frac{\gamma_1}{2}+\frac{\gamma_2}{2}+\frac{\gamma_3}{2}}  \nonumber  \\
& \times (\textcolor{OliveGreen}{y_1}(1-\textcolor{OliveGreen}{y_1}))^{J_1+\frac{\gamma_1}{2}}(\textcolor{blue}{U_1}(1-\textcolor{OliveGreen}{y_1})+\textcolor{red}{u_2} \textcolor{blue}{U_2}(1-\textcolor{OliveGreen}{y_2})\textcolor{OliveGreen}{y_1}+\textcolor{blue}{U_1} \textcolor{blue}{U_2} \textcolor{OliveGreen}{y_1} \textcolor{OliveGreen}{y_2})^{-1-J_1-J_2+\ell_1+\ell_2+\frac{\gamma_3}{2}-\frac{\gamma_2}{2}-\frac{\gamma_1}{2}}  \nonumber  \\
& \times (\textcolor{OliveGreen}{y_2}(1-\textcolor{OliveGreen}{y_2}))^{J_2+\frac{\gamma_2}{2}} (\textcolor{blue}{U_2}(1-\textcolor{OliveGreen}{y_3})+\textcolor{red}{u_6} \textcolor{blue}{U_3}(1-\textcolor{OliveGreen}{y_1})\textcolor{OliveGreen}{y_3}+\textcolor{blue}{U_2} \textcolor{blue}{U_3} \textcolor{OliveGreen}{y_1} \textcolor{OliveGreen}{y_3})^{-1-J_1-J_3+\ell_1+\ell_3+\frac{\gamma_2}{2}-\frac{\gamma_1}{2}-\frac{\gamma_3}{2}}  \nonumber \\
& \times (\textcolor{OliveGreen}{y_3}(1-\textcolor{OliveGreen}{y_3}))^{J_3+\frac{\gamma_3}{2}} (\textcolor{blue}{U_3}(1-\textcolor{OliveGreen}{y_2})+\textcolor{red}{u_4} \textcolor{blue}{U_1}(1-\textcolor{OliveGreen}{y_3})\textcolor{OliveGreen}{y_2}+\textcolor{blue}{U_1} \textcolor{blue}{U_3} \textcolor{OliveGreen}{y_2} \textcolor{OliveGreen}{y_3})^{-1-J_2-J_3+\ell_2+\ell_3+\frac{\gamma_1}{2}-\frac{\gamma_2}{2}-\frac{\gamma_3}{2}}  \nonumber  \\
& \times  \left( \textcolor{blue}{U_1}-\textcolor{red}{u_2}\textcolor{blue}{U_2} + \textcolor{blue}{U_2}(\textcolor{red}{u_2}-\textcolor{blue}{U_1})\textcolor{OliveGreen}{y_2} - \textcolor{blue}{U_1}(\textcolor{blue}{U_3}-1)\textcolor{OliveGreen}{y_3} + (\textcolor{blue}{U_2}-\textcolor{red}{u_6}\textcolor{blue}{U_3})(\textcolor{blue}{U_1} \textcolor{OliveGreen}{y_2} \textcolor{OliveGreen}{y_3} + \textcolor{red}{u_2}(1-\textcolor{OliveGreen}{y_2})\textcolor{OliveGreen}{y_3} \right))^{J_1-\ell_2-\ell_3} \nonumber \\
& \times \left( \textcolor{blue}{U_3}-\textcolor{red}{u_4}\textcolor{blue}{U_1} + \textcolor{blue}{U_1}(\textcolor{red}{u_4}-\textcolor{blue}{U_3})\textcolor{OliveGreen}{y_3} - \textcolor{blue}{U_3}(\textcolor{blue}{U_2}-1)\textcolor{OliveGreen}{y_1} + (\textcolor{blue}{U_1}-\textcolor{red}{u_2}\textcolor{blue}{U_2})(\textcolor{blue}{U_3} \textcolor{OliveGreen}{y_1} \textcolor{OliveGreen}{y_3} + \textcolor{red}{u_4}(1-\textcolor{OliveGreen}{y_3})\textcolor{OliveGreen}{y_1} \right))^{J_2-\ell_1-\ell_3}  \nonumber \\
& \times \left( \textcolor{blue}{U_2}-\textcolor{red}{u_6}\textcolor{blue}{U_3} + \textcolor{blue}{U_3}(\textcolor{red}{u_6}-\textcolor{blue}{U_2})\textcolor{OliveGreen}{y_1} - \textcolor{blue}{U_2}(\textcolor{blue}{U_1}-1)\textcolor{OliveGreen}{y_2} + (\textcolor{blue}{U_3}-\textcolor{red}{u_4}\textcolor{blue}{U_1})(\textcolor{blue}{U_2} \textcolor{OliveGreen}{y_1} \textcolor{OliveGreen}{y_2} + \textcolor{red}{u_6}(1-\textcolor{OliveGreen}{y_1})\textcolor{OliveGreen}{y_2} \right))^{J_3-\ell_1-\ell_2} \,. \nonumber
\end{align}

\section{Decomposition through Casimirs}
\label{app:Casimirs}
The snowflake light-cone blocks
\beq
\mathbb{F}_{J_1,J_2,J_3,\ell_1,\ell_2,\ell_3} \equiv \int_0^1 dy_1  \int_0^1 dy_2  \int_0^1dy_3\, \mathcal{F}
\eeq
obey three important Casimir equations:
\beqa
&&\big(\widehat{C}_{12} - \mathcal{C}_{\Delta_{J_1},J_1} \big)  \cdot
\mathbb{F}_{J_1,J_2,J_3,\ell_1,\ell_2,\ell_3} =0 \,, \la{P1}\\
&&\big(\widehat{C}_{34} - \mathcal{C}_{\Delta_{J_2},J_2} \big)  \cdot
\mathbb{F}_{J_1,J_2,J_3,\ell_1,\ell_2,\ell_3}=0\,, \la{P2}\\
&&\big(\widehat{C}_{56} - \mathcal{C}_{\Delta_{J_3},J_3} \big)  \cdot
\mathbb{F}_{J_1,J_2,J_3,\ell_1,\ell_2,\ell_3}=0,\,  \la{P3}
\eeqa
where $\mathcal{C}_{\Delta,J}=J(-2+d+J)+\Delta(\Delta-d)$ is the Casimir eigenvalue\footnote{We are in four dimensions so that $d=4$ however the leading twist expansion is known to be dimension independent since the kinematics of the OPE is governed by two dimensional light-cone plane. It is still convenient for debugging purposes to leave $d$ unevaluated in all intermediate steps and check that the $d$ dependence drops out in the end.} and  $\widehat{C}_{ij}$ represents the light-cone Casimir operator which we can obtain from 
\begin{align}
&\widehat{C}_{ij}f\left(u_1,\dots,U_3\right)=(x_{12}^2x_{34}^2x_{56}^2)^{\Delta_{\phi}}\bigg[\frac{1}{2}(L_{AB,i}+L_{AB,j})^2\bigg]\frac{1}{(x_{12}^2x_{34}^2x_{56}^2)^{\Delta_{\phi}}}f\left(u_1,\dots,U_3\right)\bigg|_{u_{2i-1}\rightarrow 0}
\end{align}
where the $L_{AB,i}$ is a generator of the conformal group and $u_{2i-1}\rightarrow 0$ stands for the leading term in this limit. It is convenient to introduce  yet another set of cross ratios given by
\beqa
&u_2=\frac{\left(1-z_2\right) \left(1-z_1 z_2 \hat{z}_3\right)}{1-z_2 z_3 \hat{z}_1},\, \ \ \ \ u_4=\frac{\left(1-z_3\right) \left(1-z_2 z_3 \hat{z}_1\right)}{1-z_1 z_3
   \hat{z}_2},\, \ \ \ \ u_6=\frac{\left(1-z_1\right) \left(1-z_1 z_3 \hat{z}_2\right)}{1-z_1 z_2 \hat{z}_3},\nonumber\\
&U_1=\frac{1-z_2}{1-z_2 z_3 \hat{z}_1},\,\, \ \ \ \ U_2=\frac{1-z_1}{1-z_1 z_2
   \hat{z}_3},\,\, \ \ \ \ U_3=\frac{1-z_3}{1-z_1 z_3 \hat{z}_2}. 
\eeqa
The first few terms of the Casimir differential operator $\widehat{C}_{12}$, in these new variables, reads
\begin{align}
& \frac{\widehat{C}_{12}}{2}=u_1^2 \left(z_1-2\right)\partial_{u_1}^2  z_1^2 z_3 \hat{z}_2^2 u_5\partial _{u_5} \partial _{\hat{z}_2}+\frac{ z_1^2 \left(z_2 \hat{z}_3-1\right){}^2}{\left(z_1-1\right)\left(z_1 z_2 \hat{z}_3-1\right)}u_3 \hat{z}_3 \partial _{u_3} \partial _{\hat{z}_3}+\dots\, .
\end{align}

We like these new variables  because of the most transparent OPE ($z_i,\hat{z}_i\rightarrow 0$) boundary conditions:
\beqa
\mathbb{F}_{J_1,J_2,J_3,\ell_1,\ell_2,\ell_3} \simeq \frac{1}{2^{\sum_{i}(J_i-l_i)}}z_1^{J_1}z_2^{J_2}z_3^{J_3}\hat{z}_{1}^{l_1}\hat{z}_2^{l_2}\hat{z}_3^{l_3}  
\eeqa
Given a perturbative data \texttt{data} (see next subsection (\ref{eq:definitionofdatatoOPE})) we can then extract any OPE data using the projections (\ref{P1},\ref{P2},\ref{P3}) as 
\beqa
\sum_{\ell_1,\ell_2,\ell_3} P_{123}^{\bullet \bullet \bullet}(J_1,J_2,J_3,\ell_1,\ell_2,\ell_3) \hat{z}_1^{\ell_1}  \hat{z}_1^{\ell_2}  \hat{z}_3^{\ell_3} = \lim_{z_{1,2,3} \to 0} \frac{1}{z_1^{J_1} z_2^{J_2} z_3^{J_3} }  \,\texttt{data}_{J_1,J_2,J_3} \nonumber
\eeqa
where 
\beq
\texttt{data}_{J_1,J_2,J_3} \equiv  \prod_{j_3<J_3} \frac{ \widehat{C}_{56} - \mathcal{C}_{\Delta_{j_3},j_3}}{ \mathcal{C}_{\Delta_{J_3},J_3}  - \mathcal{C}_{\Delta_{j_3},j_3}}  \cdot  
\prod_{j_2<J_2} \frac{ \widehat{C}_{34} - \mathcal{C}_{\Delta_{j_2},j_2}}{ \mathcal{C}_{\Delta_{J_2},J_2}  - \mathcal{C}_{\Delta_{j_2},j_2}}  \cdot 
\prod_{j_1<J_1} \frac{ \widehat{C}_{12} - \mathcal{C}_{\Delta_{j_1},j_1}}{ \mathcal{C}_{\Delta_{J_1},J_1}  - \mathcal{C}_{\Delta_{j_1},j_1}}  \, \cdot \, \texttt{data}\label{CasimirProjection}
\eeq
 is the perturbative data with spins smaller than $J_1,J_2,J_3$ projected out. Every time we act with Casimir on the conformal block we get back the block times its Casimir eigenvalue. The denominator in (\ref{CasimirProjection}) is chosen such that the coefficient multiplying power of $z_i^{i}\hat{z}_i^{\ell_i}$ is the OPE coefficient.

\begin{figure}[t]
\includegraphics[width=\textwidth]{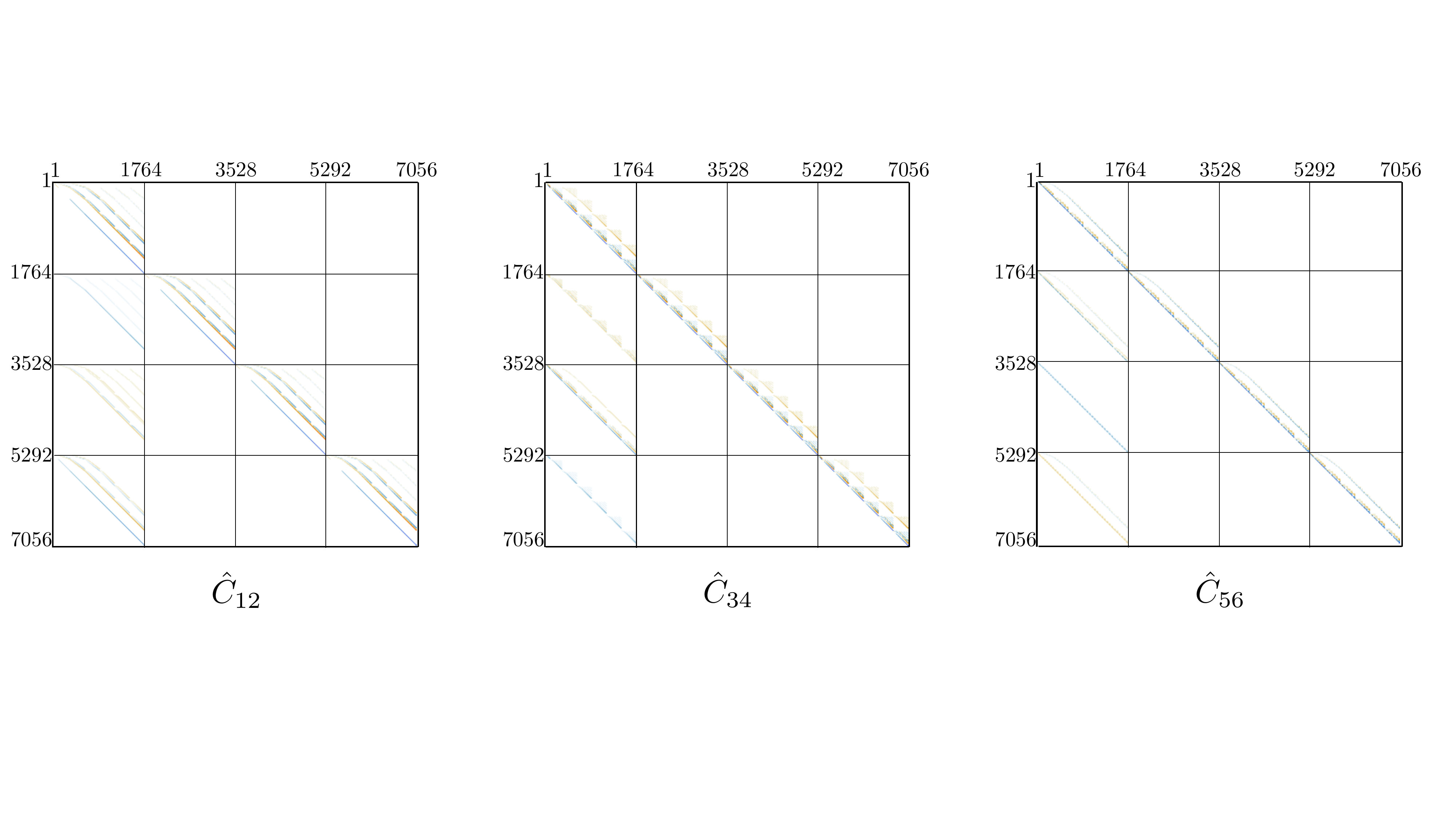}
\vspace{-3cm}
\caption{\small The six point function in the snowflake OPE limit ($z_i$, $\hat{z}_i$ $\rightarrow 0$) can be expressed as a sum of monomials  $\prod_i z_i^{a_i} \hat{z}_i^{l_i}$ which in perturbation theory can be dressed by $\log(u_i)$. The action of the Casimir operators in this base of dressed monomials can be represented as sparse matrices which are plotted here for $a_i\leq6$, $l_i\leq2$. The basis is restricted to $l_1 + l_2 \leq a_3$ and similar for the other $a_i$ as discused in the main text. At one loop, the basis can be divided into terms with no $\log$s, or a single $\log(u_1)$, $\log(u_2)$ or $\log(u_3)$, leading to the $4\times4$ block structure. 
The Casimir can remove $log$s but never generate them, justifying the absence of off diagonal terms in the last three columns.}
\label{matrixcasimir}
\end{figure}

This way of extracting is very convenient. The data itself is organized in a simple manner, for given power of $z_i^{a_i}$ the $\hat{z}_i^{l_i}$ powers are such that $l_1+l_2\le J_3$ is satisfied and analogously for the others $l_i$ and at one loop these powers can be dressed by $\ln u_i$. An  efficient way to do this extraction is transform the Casimir into a matrix that acts on a vector space created powers of $z_i, \hat{z}_i$ and $\ln u_i$ which then makes (\ref{CasimirProjection}) into a matrix multiplication problem, see figure \ref{matrixcasimir}. Obviously we need to consider finite dimensional vector spaces so we take a cut off ($\Lambda_i, \hat{\Lambda}_i$) in the powers of $z_{i}, \hat{z}_i$. This allows to extract OPE coefficients up to spin $J_i=\Lambda_i, \ell_i=\hat{\Lambda}_i$.  With this we manage to extract around three hundred thousand OPE coefficients at one loop order which we will analyze in the next section. 
\tocless\subsection{Data}
\label{appendix1Loopdata}
Perturbative results for three point functions with more than one spinning operator are considerably more complicated to compute than with just one operator with spin.  While three point functions with one spinning operator have been computed, in $\mathcal{N}=4$ SYM, up to three loops \cite{Eden:2012rr},   three point functions with two spinning operators have only been computed up to one loop \footnote{There are some results for low spinning operators up to two loops \cite{Marco}. }. In the following we will compute these three point functions at one loop for both one, two and three spinning operators by doing conformal block decomposition decomposition of a one loop six point function. 

The starting point is the six point function of $20'$ operators
\begin{align}
\mathcal{O}_2(x)= y^Iy^J\textrm{Tr}(\phi^{I}(x)\phi^{J}),\,  \ \  \ y^2=0,\, \, \ \ \   I=1,\dots 6
\end{align}
which has been computed at one loop in \cite{Drukker:2008pi} and it is expressed as a simple linear combination of one loop four point integrals $I_{i_1i_2i_3i_4}$(see eq. (32,60-61) of \cite{Drukker:2008pi} for the precise definition of the six point function)
\begin{align}
I_{i_1i_2i_3i_4} = \int \frac{d^4x_0}{x_{i_10}^2x_{i_20}^2x_{i_30}^2x_{i_40}^2}
\end{align}
which can be easily computed in terms of cross ratios.

%

We are interested in obtaining three point functions in the $[0,2,0]$ representation of $SO(6)$ in the OPE. This can be achieved by projecting appropriately the null polarization vectors $y$ into this particular representation (see appendix B of \cite{Goncalves:2019znr} for details). Then we take the light-cone limits $x_{12}^2,x_{34}^2,x_{56}^2\rightarrow 0$ to focus on leading twist operators. At the end of this procedure we arrive at the following (tree level plus one loop) expression
\begin{align}
\texttt{data} &= (x_{12}^2x_{34}^2x_{56}^2)^2\langle \mathcal{O}_2(x_1)\dots \mathcal{O}_2(x_6)\rangle\bigg|_{\texttt{projection on $[0,2,0]$}\atop {x_{12}^2,x_{34}^2,x_{56}^2\rightarrow 0}  }.\label{eq:definitionofdatatoOPE}
\\
&= \prod_{i=1}^3\frac{\textcolor{OliveGreen}{u_{2i-1}}}{u_{2i}U_{i}}\bigg[\frac{3u_2u_4\textcolor{blue}{U_1}\textcolor{blue}{U_2}(1+u_6)+\textcolor{blue}{U_1}\textcolor{blue}{U_2}\textcolor{blue}{U_3}(1+u_2u_4u_6)}{24}+\nonumber\\
&-\frac{\textcolor{red}{\lambda}}{4}\bigg(\ln \textcolor{OliveGreen}{u_1}\big[\textcolor{blue}{U_3}\ln u_2 (u_4u_6(\textcolor{blue}{U_1}-u_2\textcolor{blue}{U_2})+\textcolor{blue}{U_2}(\textcolor{blue}{U_1}-u_2u_6))\nonumber\\
&+\textcolor{blue}{U_2}\ln u_6(u_2(u_4(1-u_6)\textcolor{blue}{U_1}-u_6\textcolor{blue}{U_3})+\textcolor{blue}{U_1}\textcolor{blue}{U_3})+u_4u_6\textcolor{blue}{U_3}\ln \textcolor{blue}{U_2} ((1+u_2)\textcolor{blue}{U_1}-u_2(1+\textcolor{blue}{U_1})\textcolor{blue}{U_2})\nonumber\\
&+\textcolor{blue}{U_3}\ln \textcolor{blue}{U_1}(\textcolor{blue}{U_2}(u_2u_6-\textcolor{blue}{U_1})+u_4u_6(u_2\textcolor{blue}{U_2}-\textcolor{blue}{U_1}))\big]+\ln u_2\big[u_2u_6\textcolor{blue}{U_3}\ln \textcolor{blue}{U_3} (\textcolor{blue}{U_2}-u_4\textcolor{blue}{U_1})\nonumber\\
&+\textcolor{blue}{U_1}\ln \textcolor{blue}{U_2}(u_2u_4(\textcolor{blue}{U_2}+u_6\textcolor{blue}{U_3})-\textcolor{blue}{U_3}(\textcolor{blue}{U_2}+u_4u_6))\big]+\textcolor{blue}{U_2}\ln \textcolor{blue}{U_1}\ln \textcolor{blue}{U_2}(\textcolor{blue}{U_1}\textcolor{blue}{U_3}+u_2u_6(u_4\textcolor{blue}{U_1}-(1+u_4)\textcolor{blue}{U_3}))\nonumber\\
&+u_6(u_4\textcolor{blue}{U_1}\textcolor{blue}{U_3}+u_2(\textcolor{blue}{U_2}\textcolor{blue}{U_3}-u_4\textcolor{blue}{U_1}(\textcolor{blue}{U_2}+\textcolor{blue}{U_3})))\texttt{Li}_2(1-u_2) \nonumber\\
&+\textcolor{blue}{U_2}(\textcolor{blue}{U_1}\textcolor{blue}{U_3}-u_2u_4(\textcolor{blue}{U_1}+u_6\textcolor{blue}{U_3}(\textcolor{blue}{U_1}-1)))\texttt{Li}_2(1-\textcolor{blue}{U_1})\nonumber\\
&+\big(\textcolor{blue}{U_1}\textcolor{blue}{U_3}(u_4u_6+\textcolor{blue}{U_2})+u_2(u_4\textcolor{blue}{U_1}\textcolor{blue}{U_2}(u_6-1)+u_6\textcolor{blue}{U_3}(\textcolor{blue}{U_2}(u_4(\textcolor{blue}{U_1}-1)-1)-\textcolor{blue}{U_1}u_4))\big)\texttt{Li}_2\left(1-\frac{u_2}{\textcolor{blue}{U_1}}\right)\nonumber\\
&+\textcolor{blue}{U_2}(\textcolor{blue}{U_1}\textcolor{blue}{U_3}-u_2u_4(\textcolor{blue}{U_1}+u_6\textcolor{blue}{U_3}(\textcolor{blue}{U_1}-1)))\texttt{Li}_2\left(1-\frac{u_2u_4}{\textcolor{blue}{U_3}}\right)\nonumber\\
&+u_4(u_6\textcolor{blue}{U_1}\textcolor{blue}{U_3}-u_2\textcolor{blue}{U_2}(u_6\textcolor{blue}{U_3}+\textcolor{blue}{U_1}(u_6-1)))\texttt{Li}_2\left(1-\frac{u_2\textcolor{blue}{U_2}}{\textcolor{blue}{U_1}}\right)\bigg)+\texttt{two permutations}\bigg]\nonumber\\
&=\prod_{i=1}^3\textcolor{OliveGreen}{u_{2i-1}}\bigg[1+\frac{\sum_{i}(z_{i}+z_i^2)}{2}+\frac{z_1z_2+z_1z_3+z_2z_3}{4}-\frac{z_2z_3\hat{z}_1+z_1z_3\hat{z}_2+z_1z_2\hat{z}_3}{4}+\nonumber\\
&+\textcolor{red}{\lambda} \big(z_1^2(\ln \textcolor{OliveGreen}{u_1}-2 )+z_2^2(\ln \textcolor{OliveGreen}{u_3}-2 )+z_3^2(\ln \textcolor{OliveGreen}{u_5}-2)\big)+\dots\bigg]
\end{align}
where the two permutations are just given by $u_i,U_i\rightarrow u_{i+2},U_{i+2};u_{i+4},U_{i+4}$ and where the dots in the last line stand for higher order powers in $z_i$. 
This is precisely the object that enters in (\ref{CasimirProjection}) and that can be used to extract OPE coefficients of three spinning operators at one loop. The last line is transformed into a vector of the monomials and $\log$s which is then acted by the Casimir matrix of figure \ref{matrixcasimir} to efficiently extract all the needed OPE data.

%
%


Through this method we extracted over three hundred thousand OPE coefficients. Both the notebook used for extraction as well as a sample of the result are presented in the attached \texttt{Mathematica} notebooks. Our goal now is to write such data as an expression analytically both in spins and polarizations. In the next three sections we display the structure constant for one, two and three spinning operators.

\section{One Loop Explorations} 
\label{appendix1Loop}
\subsection{One spinning operator}
The structure constants for one spinning operator have been studied extensively in the literature. The first non trivial computation of these three point functions in $\mathcal{N}=4$ was done at one loop level in \cite{Dolan:2004iy} via conformal block decomposition of a four point function. It is the most efficient way of computing these three point function. Using this method the structure constants with one spinning operator have been computed up to three loops for any spin in \cite{Eden:2012rr} and is known up to five loops for spin two \cite{Georgoudis:2017meq}. 

The one loop structure constant simply reads
\begin{equation}
\hat{C}^{\circ\circ\bullet}_{\texttt{1-loop}} = 4S_1(J)^2-4S_1(J)S_1(2J)-2S_2(J)
\label{cOneSpin}
\end{equation}
where $S_i$ are the harmonic functions.

To set the coupling convention let us quote here the dimension of these operators
\begin{equation}
\Delta_J=2+J+8\lambda S_1(J)+O(\lambda^2)
\end{equation}
so that from large spin we read
\begin{equation}
f=8\lambda+O(\lambda^2)\,, \quad \text{and} \quad g=8\lambda \gamma_E+O(\lambda^2)\,.
\end{equation}

\subsection{Two spinning operators}
\label{twospinning}
Here we complete Bianchi's computation \cite{Marco} for the one loop structure constant
\begin{align}
\hat{C}^{\circ\bullet\bullet}_{\texttt{1-loop}} & = 4\left(S_1(J_1)+S_1(J_2)\right)\sum_{i=1}^{\ell}\frac{\binom{J_1+J_2+1}{i}\binom{\ell}{i}}{i \binom{J_1}{i}\binom{J_2}{i}}-4 \frac{(J_1+1)(J_2+1)}{J_1+J_2+2}\sum_{i=1}^\ell\frac{\binom{J_1+J_2+2}{i}a_i^\ell}{i \binom{J_1}{i}\binom{J_2}{i}} + \nonumber \\
& +\sum_{i=1}^2 4S_1(J_i)^2-4S_1(J_i)S_1(2J_i)-2S_2(J_i)\label{cTwoSpins}
\end{align}
where  the constants $a_i^\ell$  were only known in special limiting cases. They are given by 
\begin{equation}
a_i^\ell= \frac{(-1)^i}{2} \left( -S_1(\ell)^2 - S_2(\ell) + 
2\sum_{k=2}^{i} \frac{(-1)^k\binom{l}{k-1}}{k-1}\left(S_1(k-2)+S_1(\ell)\right)\right) \,.
\label{as}
\end{equation}
For $i=1$ and for $i=\ell$ it simplifies to the two previously known special cases (4.5) and (4.6) in~\cite{Marco}.
This one loop structure constant can also be rewritten as (\ref{BiBianchi}), where
\begin{equation}
    \mathbb{B}(j,l)=(-1)^l\binom{j}{l}\,.
    \label{BoldB}
\end{equation}
Having the full expression for the structure constant at one loop in the form of (\ref{BiBianchi}) allow us to study the physics of this correlator at large spin and polarizations, with $\frac{\ell}{J_i}$ fixed.  As discussed in section \ref{oneloop}, this serves as a toy model for the large spin limit of three point functions of spinning operators and its relation to the null hexagonal Wilson loops. 

\begin{align}
    \hat{C}^{\circ\bullet\bullet}_{\texttt{1-loop}}& =8 S_1(J_1)^2 - 4 S_1(J_1) S_1(2 J_1) + 8 S_1(J_1) S_1(J_2) + 8 S_1(J_2)^2 -  4 S_1(J_2) S_1(2 J_2)+\nonumber\\
    & - 4 S_1(J_1) S_1(J_1 - \ell) - 4 S_1(J_2) S_1(J_1 - \ell) -  4 S_1(J_1) S_1(J_2 - \ell) - 4 S_1(J_2) S_1(J_2 - \ell)+\nonumber\\
    & - 4 S_1(J_1) S_1(\ell) -  4 S_1(J_2) S_1(\ell) + 4 S_1(J_1 - \ell) S_1(\ell) + 4 S_1(J_2 - \ell) S_1(\ell) - 
 2 S_1(\ell)^2+\nonumber\\
 & - 2 S_2(J_1) - 2 S_2(J_2) -  2 S_2(\ell)-4\sum_{p=1}^{\infty}\Bigg(\left(\frac{\mathbb{B}(J_1,\ell-p)}{\mathbb{B}(J_1,\ell)} + \frac{\mathbb{B}(J_2,\ell-p)}{\mathbb{B}(J_2,\ell)}\right)\frac{S_1(p-1)}{p}+\nonumber\\
 & +4(S_1(J_1)+S_1(J_2)-S_1(\ell))\frac{\mathbb{B}(J_1,\ell-p)}{\mathbb{B}(J_1,\ell)}\frac{\mathbb{B}(J_2,\ell-p)}{\mathbb{B}(J_2,\ell)}\frac{1}{p} -4 \frac{\mathbb{B}(J_1,\ell-p)}{\mathbb{B}(J_1,\ell)} \times \nonumber\\
 &\times \frac{\mathbb{B}(J_2,\ell-p)}{\mathbb{B}(J_2,\ell)}\frac{S_1(p)+S_1(\ell-p)-S_1(\ell)}{p}  +\sum_{q=1}^{p-1}\frac{\mathbb{B}(J_1,\ell-p)}{\mathbb{B}(J_1,\ell)}\frac{\mathbb{B}(J_2,\ell-q)}{\mathbb{B}(J_2,\ell)}\frac{1}{(p-q)p} \nonumber\\ 
 &+ \sum_{q=p+1}^{\infty}\frac{\mathbb{B}(J_1,\ell-p)}{\mathbb{B}(J_1,\ell)}\frac{\mathbb{B}(J_2,\ell-q)}{\mathbb{B}(J_2,\ell)}\frac{1}{(q-p)q} \Bigg)\label{BiBianchi}.
\end{align}

There are two regimes of interest, depending on whether $\ell $ is before or after $\ell_* \equiv \frac{J_1 J_2}{\left(J_1 + J_2\right)}$.\footnote{ Note that the tree level three point function, given by
\begin{equation*}
\nonumber C_{\text{tree}}^{\bullet \bullet \circ}=\frac{\Gamma(J_1+1) \Gamma(J_2 +1)}{\sqrt{\Gamma(2J_1+1)\Gamma(2J_2+1)}\Gamma(J_1-\ell +1)\Gamma(J_2 - \ell +1)\Gamma(\ell +1)},
\end{equation*}
approaches a gaussian centered at $\ell_*$ in the large spin limit. } Before $\ell_*$ we obtain the order one result 
\begin{align}
\la{beforeSP}  \hat{C}^{\bullet \bullet \circ}_{\texttt{1-loop}} \xrightarrow{J_1,J_2,\ell \rightarrow \infty}& -\pi^2 - 4\log(2) \log\left(J_1 J_2 e^{2\gamma_E} \right)  - 2  \log^2\left(\ell e^{\gamma_E}\right) \\&
- 2 \log\Big(1 - \frac{\ell}{J_1} -\frac{ \ell}{J_2}\Big)  \log\Big(\frac{e^{2\gamma_E}J_1^2 J_2^2}{\ell^2}\big(1 - \frac{\ell}{J_1} -\frac{ \ell}{J_2}\big) \Big)\nonumber
\end{align}
while after $\ell_*$ we obtain the exponentially large expression \begin{align}
\la{afterSP}  \hat{C}^{\bullet \bullet \circ}_{\texttt{1-loop}}  \xrightarrow{J_1,J_2,\ell \rightarrow \infty} & \frac{J_1^{-2 J_1-1} J_2^{-2 J_2-1} \left(J_1+J_2\right){}^{J_1+J_2+\frac{3}{2}} \ell ^{2 \ell +1} \nonumber \left(J_1-\ell \right){}^{J_1-\ell +\frac{1}{2}} \left(J_2-\ell  \right){}^{J_2-\ell +\frac{1}{2}}}{J_1 J_2-J_1 \ell -J_2 \ell} \times\\& 
\times 4 \sqrt{2 \pi} \log\left(\frac{\ell}{J_1} + \frac{\ell}{J_2} -1 \right).
\end{align}
Note that the singular point $\ell_*$ is reminiscent of the singularities encountered in the $U(\ell)$ map, equation (\ref{Ulmap}). The $\log$ singularities at $\ell_*$ in (\ref{beforeSP}) should therefore be compared with the $\log$ singularities of the hexagonal Wilson loop at $U_i=0$, see (\ref{repAB}). This justifies treating $C^{\bullet \bullet \circ}$ as a toy model for the transition in behaviour of $C^{\bullet \bullet \bullet}$ from the space-like region to the time-like region in section \ref{oneloop}.

Lets comment on how these results can be derived. The only non-trivial pieces in (\ref{BiBianchi}) are the sums, of which there are two types, that with a single ratio of binomials, as in the third line, and those with a product of ratio of binomials, as in the last three lines. First we analyze the latter. 

Binomials $\binom{j}{l}$, in the large $j$ limit with $\frac{\ell}{j}$ fixed approach gaussians with mean $j/2$ and standard deviation $\sqrt{j}/2$. The product of binomials in the denominators therefore approach gaussians\footnote{The product of gaussians is a gaussian.} with mean $\ell_*= \frac{J_1 J_2}{J_1 + J_2}$ and standard deviation $O(\sqrt{J})$. Similar happens in the numerators only that arguments are shifted. The sum indices $p$ and $q$ being positive, we see that if $\ell < \ell_*$ so that the denominator is evaluated to the left of the maximum, large $p$ and $q$ are exponentially suppressed relative to $p$, $q$ of order one. In this limit, the sums are easily evaluated. For example, in the fifth line we have the simplification
\beq 
\sum_{p=1}^{l}\left(\frac{\mathbb{B}(J_1,\ell-p)}{\mathbb{B}(J_1,\ell)}\frac{\mathbb{B}(J_2,\ell-p)}{\mathbb{B}(J_2,\ell)}\frac{1}{p}  \rightarrow
\left(\frac{\ell^2}{(J_1-\ell)(J_2-l)}\right)^p \frac{1}{p}\right) = -\log \left(\frac{\ell(J_1 + J_2) -J_1 J_2}{(J_1-\ell)(J_2-\ell)}\right). \la{binomialsum}
\eeq On the other hand, if $\ell > \ell_*$, provided $p$ and $q$ are $O(J)$, we can tune the indices so that the numerator sits at the top of the gaussian. The sums are therefore evaluated by saddle point and we obtain the exponential expression in (\ref{afterSP}).

 For the summand with a single ratio of binomials, for similar reasons, $p$ of order one dominates when $\ell < \ell_*$ and $\log$s are generated. However, for this term there are no exponential contributions when $\ell > \ell_*$ due to the oscillating phase in (\ref{BoldB}). Therefore this term can be neglected in the $\ell > \ell_*$ regime.

Finally, we note a sum rule for the one-loop sum of structure constants, valid at finite $J_1$ and $J_2$:
\beq
\label{sum2spin}
\sum_\ell C^{\bullet \bullet \circ}_{\texttt{1-loop}} =   4 \left(S_1(J_1) + S_1(J_2)\right) S_1(J_1 + J_2) - \sum_{i=1}^2 4 S_1(J_i)S_1(2 J_i)+ 2 S_2(J_i)\,,
\eeq
note that this sums is for the full one loop correction to the structure constants ($C=C_{\texttt{tree}}+C_{\texttt{1-loop}}$)  and \textit{not} the correction normalized by tree level ($\hat{C}$).


\subsection{Three spinning operators} \label{appe}
For three spinning operators we were not able to find an expression analogous (\ref{cTwoSpins}) or (\ref{BiBianchi}), which is analytical both in spin and polarizations. In this section we present a general expression for the one-loop structure constant in terms of unknown coefficients that we could not fix entirely. However, when considering small polarizations (where we have abundant perturbative data) we were able to fix such coefficients and arrive in an analytic expression for the structure constant in terms of the spins $J_i$, such as (\ref{data123}).

We start by parametrizing the one-loop structure constant as the sum of three terms
\begin{equation}
    \hat{C}^{J_1,J_2,J_3}_{\ell_1,\ell_2,\ell_3}= \mathcal{X}^{J_1,J_2,J_3}+\mathcal{Y}^{J_1,J_2,J_3}_{\ell_1,\ell_2,\ell_3}+\mathcal{Z}^{J_1,J_2,J_3}_{\ell_1,\ell_2,\ell_3}
    \label{C3spins}
\end{equation} 

The first term of (\ref{C3spins}) is a generalization of the one spin structure constant (\ref{cOneSpin}), and it is simply given by
\begin{align}
\mathcal{X}^{J_1,J_2,J_3} &= 4S_1(J_1)^2-4S_1(J_1)S_1(2J_1)-2S_2(J_1)
+4S_1(J_2)^2-4S_1(J_2)S_1(2J_2)+\nonumber\\
& -2S_2(J_2)+ 4S_1(J_3)^2-4S_1(J_3)S_1(2J_3)-2S_2(J_3)
\label{trivialPart}
\end{align}
this term gives the one-loop structure constants for vanishing polarizations ($\ell_i=0$).

The second term of (\ref{C3spins}) is inspired in the two spin structure constants (\ref{cTwoSpins}), which reads
\begin{align}
\mathcal{Y}^{J_1,J_2,J_3}_{\ell_1,\ell_2,\ell_3} &= 4\left(S_1(J_1)+S_1(J_2)\right)\sum_{i=1}^{\ell_3}\left(\frac{\binom{J_1+J_2+1}{i}\binom{\ell_3}{i}}{i \binom{J_1}{i}\binom{J_2}{i}}\right)-4\frac{(J_1+1)(J_2+1)}{J_1+J_2+2}\sum_{i=1}^{\ell_3}\left(\frac{\binom{J_1+J_2+2}{i}a_i^{\ell_3}}{i \binom{J_1}{i}\binom{J_2}{i}} \right)+\nonumber\\
& +4\left(S_1(J_1)+S_1(J_3)\right)\sum_{i=1}^{\ell_2}\left(\frac{\binom{J_1+J_3+1}{i}\binom{\ell_2}{i}}{i \binom{J_1}{i}\binom{J_3}{i}}\right)-4\frac{(J_1+1)(J_3+1)}{J_1+J_3+2}\sum_{i=1}^{\ell_2}\left(\frac{\binom{J_1+J_3+2}{i}a_i^{\ell_2}}{i \binom{J_1}{i}\binom{J_3}{i}} \right)+\nonumber \\
& +4\left(S_1(J_2)+S_1(J_3)\right)\sum_{i=1}^{\ell_1}\left(\frac{\binom{J_2+J_3+1}{i}\binom{\ell_1}{i}}{i \binom{J_2}{i}\binom{J_3}{i}}\right)-4\frac{(J_2+1)(J_3+1)}{J_2+J_3+2}\sum_{i=1}^{\ell_1}\left(\frac{\binom{J_2+J_3+2}{i}a_i^{\ell_1}}{i \binom{J_2}{i}\binom{J_3}{i}} \right)+\nonumber\\
&+ 4\sum_{i=1}^3 S_1(\ell_i)S_1(j_i).
\end{align}
This expression is a simple symmetrization of (\ref{cTwoSpins}), except the last line, which does not exist in the context of two spinning operators. When combined with $\mathcal{X}^{J_1,J_2,J_3}$ it gives the structure constants with two vanishing polarizations ($\hat{C}^{J_1,J_2,J_3}_{0,0,\ell}$).

Through lengthy explorations of the extracted one loop data and inspired by \cite{Marco} we were able to parametrize any one-loop structure constant via the following ansatz
\begin{equation}
    \mathcal{Z}^{J_1,J_2,J_3}_{\ell_1,\ell_2,\ell_3} = q\left(J_1,J_2,J_3,\ell_1,\ell_2\right)+q\left(J_2,J_3,J_1,\ell_2,\ell_3\right)+q\left(J_3,J_1,J_2,\ell_3,\ell_1\right)
\end{equation}
where
\begin{align}
    \frac{q\left(j_1,j_2,j_3,l_1,l_2\right)}{(1+j_1)(1+j_2)} & =  (1+j_3)\sum_{n=1}^{l_2}\sum_{m=1}^{l_1}\sum_{p=1}^{l_1+l_2}\beta^{n,m,p}_{l_1,l_2}\frac{\mathbb{B}(j_1,l_2-n)}{\mathbb{B}(j_1,l_2)}\frac{\mathbb{B}(j_2,l_1-m)}{\mathbb{B}(j_2,l_1)}\frac{\mathbb{B}(j_3,l_1+l_2-p)}{\mathbb{B}(j_3,l_1+l_2)}+\nonumber\\
    & \sum_{n=1}^{l_2}\sum_{m=1}^{l_1}\sum_{p=n+m}^{l_1+l_2}\gamma^{n,m,p}_{l_1,l_2}\frac{\mathbb{B}(j_1,l_2-n)}{\mathbb{B}(j_1,l_2)}\frac{\mathbb{B}(j_2,l_1-m)}{\mathbb{B}(j_2,l_1)}\frac{\mathbb{B}(j_3,l_1+l_2-p)}{\mathbb{B}(j_3,l_1+l_2)}S_1(j_3)    \label{qAnsatz}
\end{align}
with $\mathbb{B}$, defined in (\ref{BoldB}) and being $\beta^{n,m,p}_{l_i,l_j}$ and $\gamma^{n,m,p}_{l_i,l_j}$ unknown coefficients. We were able to fix all $\gamma^{n,m,p}_{l_i,l_j}$ coefficients and a large portion of the $\beta^{n,m,p}_{l_i,l_j}$, by comparing this ansatz with perturbative data and large spin expansions.

By matching with the extracted perturbative data we were able to fix all the $\gamma^{n,m,p}_{l_i,l_j}$ coefficients for polarizations up to $\ell_{\max}=5$ and all the $\beta^{n,m,p}_{l_1,l_2}$ for polarizations up to $\ell_{\max}=3$, being expression (\ref{data123}) below an explicit example with polarizations $\{\ell_1=1,\ell_2=2,\ell_3=3\}$.
\begingroup\makeatletter\def\f@size{10}\check@mathfonts
\def\maketag@@@#1{\hbox{\m@th\large\normalfont#1}}%
\begin{align}
\hat{C}^{J_1,J_2,J_3}_{1,2,3} &=\mathcal{X}^{J_1,J_2,J_3} + \mathcal{Y}^{J_1,J_2,J_3}_{1,2,3} + \Bigg(\frac{(1+J_2)(1+J_3)S_1(J_1)}{\mathbb{B}(J_2,3)\mathbb{B}(J_3,2)\mathbb{B}(J_1,5)}\Bigg)\Bigg(\frac{4}{5}-\frac{8J_2}{15}+\frac{13J_1J_2}{30}+ \nonumber\\
&-\frac{J_1^2J_2}{30}+\frac{2J_2^2}{15}-\frac{2J_1J_2^2}{15}+\frac{J_1^2J_2^2}{30}-\frac{2J_3}{5}+\frac{J_1 J_3}{5}+\frac{4J_2J_3}{15}-\frac{31 J_1 J_2}{60}+\frac{J_1^2J_2J_3}{5}+\nonumber\\
& \frac{J_1^3 J_2 J_3}{60} -\frac{J_2^2J_3}{15} + \frac{J_1 J_2^2 J_3}{6}-\frac{J_1^2J_2^2J_3}{10}+\frac{J_1^3 J_2^2 J_3}{60}\Bigg)+ \nonumber\\
&+\Bigg(\frac{(1+J_1)(1+J_3)S_1(J_2)}{\mathbb{B}(J_1,3)\mathbb{B}(J_3,1)\mathbb{B}(J_2,4)}\Bigg)\Bigg(1 -\frac{2J_1}{3}+ \frac{J_1^2}{6} + \frac{3 J_1 J_2}{4} - \frac{J_1^2 J_2}{4} - \frac{J_1 J_2^2}{12} + \frac{J_1^2 J_2^2}{12}\Bigg)+\nonumber\\
&+\Bigg(\frac{(1+J_1)(1+J_2)S_1(J_3)}{\mathbb{B}(J_1,2)\mathbb{B}(J_2,1)\mathbb{B}(J_3,3)}\Bigg)\Bigg(\frac{4}{3}-\frac{2J_1}{3}+\frac{2J_1J_3}{3}\Bigg)+\nonumber\\
&+\Bigg(\frac{(1+J_1)(1+J_2)(1+J_3)}{\mathbb{B}(J_1,5)\mathbb{B}(J_2,4)\mathbb{B}(J_3,3)}\Bigg)\Bigg(-\frac{17}{60}-\frac{29 J_1}{720} + \frac{71 J_1^2}{864}- \frac{61 J_1^3}{2160} + \frac{13 J_1^4}{4320}+\nonumber\\
&+\frac{13 J_2}{120} + \frac{623 J_1 J_2}{4320} - \frac{5269 J_1^2 J_2}{25920} + \frac{877 J_1^3 J_2}{12960} - \frac{187 J_1^4 J_2}{25920}-\frac{7 J_2^2}{270} - \frac{41 J_1 J_2^2}{405} + \frac{859 J_1^2 J_2^2}{6480}+\nonumber\\
&- \frac{1177 J_1^3 J_2^2}{25920}+\frac{25 J_1^4 J_2^2}{5184} + \frac{J_2^3}{270} + \frac{119 J_1 J_2^3}{6480} - \frac{103 J_1^2 J_2^3}{4320} + \frac{227 J_1^3 J_2^3}{25920} -\frac{5 J_1^4 J_2^3}{5184} +\frac{29 J_3}{120} + \nonumber \\
&+\frac{167 J_1 J_3}{1440} - \frac{173 J_1^2 J_3}{960}+\frac{83 J_1^3 J_3}{1440} - \frac{17 J_1^4 J_3}{2880} + \frac{37 J_2 J_3}{240} - \frac{153 J_1 J_2 J_3}{320} + \frac{2393 J_1^2 J_2 J_3}{5760}+\nonumber\\
&- \frac{569 J_1^3 J_2 J_3}{4320} + \frac{239 J_1^4 J_2 J_3}{17280}-\frac{73 J_2^2 J_3}{540}+\frac{215 J_1 J_2^2 J_3}{648} - \frac{1399 J_1^2 J_2^2 J_3}{5184} + \frac{1103 J_1^3 J_2^2 J_3}{12960}+\nonumber\\
& - \frac{47 J_1^4 J_2^2 J_3}{5184} + \frac{13 J_2^3 J_3}{540} - \frac{49 J_1 J_2^3 J_3}{810} + \frac{1277 J_1^2 J_2^3 J_3}{25920}-\frac{83 J_1^3 J_2^3 J_3}{5184}+\frac{23 J_1^4 J_2^3 J_3}{12960} - \frac{J_3^2}{15}+\nonumber\\
&- \frac{11 J_1 J_3^2}{288} + \frac{125 J_1^2 J_3^2}{1728} - \frac{209 J_1^3 J_3^2}{8640} + \frac{11 J_1^4 J_3^2}{4320} - \frac{53 J_2 J_3^2}{720} + \frac{5 J_1 J_2 J_3^2}{27 } - \frac{4199 J_1^2 J_2 J_3^2}{25920}\nonumber\\
&+\frac{173 J_1^3 J_2 J_3^2}{3240} - \frac{149 J_1^4 J_2 J_3^2}{25920} + \frac{31 J_2^2 J_3^2}{540} - \frac{53 J_1 J_2^2 J_3^2}{405} + \frac{1807 J_1^2 J_2^2 J_3^2}{17280}-\frac{3487 J_1^3 J_2^2 J_3^2}{103680}+\nonumber\\
&+ \frac{127 J_1^4 J_2^2 J_3^2}{34560}-\frac{11 J_2^3 J_3^2}{1080} + \frac{623 J_1 J_2^3 J_3^2}{25920} - \frac{31 J_1^2 J_2^3 J_3^2}{1620} + \frac{43 J_1^3 J_2^3 J_3^2}{6912} - \frac{73 J_1^4 J_2^3 J_3^2}{103680}\Bigg)\;.\label{data123}
\end{align}\endgroup

We now consider the large spin expansion, i.e. $J_i\to\infty$ with $\ell_i$ finite. It is easy to expand the ansatz above (\ref{qAnsatz}) up to some arbitrary order $\Lambda$. The ratio of binomials has a simple large spin limit, which allow us to truncate the sums up to $\Lambda$. This means we can trade our knowledge of knowing the $\gamma^{n,m,p}_{l_i,l_j}$ and $\beta^{n,m,p}_{l_i,l_j}$ up to some $\ell_{\max}$ to knowing the large spin expansion up to some order $\mathcal{O}(1/J^{\ell_{\max}})$.

Furthermore, in this large spin limit is easy to disentangle the terms that are associated with the coefficients $\gamma^{n,m,p}_{l_i,l_j}$ and $\beta^{n,m,p}_{l_1,l_2}$, since the first one multiplies $S_1(J_i)$ and will come together with a $\ln{J_i}$ factor. Therefore, using our perturbative data we can write the large spin expansion for the one-loop structure constant for arbitrary polarizations, up to order $\mathcal{O}(1/J^{5})$ for the terms with logs and up to order $\mathcal{O}(1/J^{3})$ for the rest (which are simple polynomials in polarizations and harmonic numbers), for example at order $\mathcal{O}(1/J^{1})$ it reads
\begin{align}
    \hat{C}^{J_1,J_2,J_3}_{\ell_1,\ell_2,\ell_3}&=-\pi^2 - 2 S_1(\ell_1)^2 - 2 S_1(\ell_1)S_1(\ell_2) - 2 S_1(\ell_2)^2 - 2 S_1(\ell_1) S_1(\ell_3) - 2 S_1(\ell_2) S_1(\ell_3)+ \nonumber\\
    & - 2 S_1(\ell_3)^2- 2 S_2(\ell_1) -  2 S_2(\ell_2) - 2 S_2(\ell_3)+4(\ln{J_1}+\gamma_E)(S_1(\ell_1)-\ln{2})+\nonumber\\
    &+4(\ln{J_2}+\gamma_E)(S_1(\ell_2)-\ln{2}+4(\ln{J_3}+\gamma_E)(S_1(\ell_3)-\ln{2})+\frac{2}{J_1} + \frac{2}{J_2} + \frac{2}{J_3}+\nonumber\\
    & - \frac{2\ln{2}}{J_1} - \frac{2\ln{2}}{J_2} - \frac{2\ln{2}}{J_3}+(\ln{J_1}+\gamma_E)\Big(\frac{1}{J_1}+\frac{4\ell_2}{J_1}+\frac{4\ell_2}{J_3}+\frac{4\ell_3}{J_1}+\frac{4\ell_3}{J_2}\Big)+\nonumber\\ 
    &+(\ln{J_2}+\gamma_E)\Big(\frac{1}{J_2}+\frac{4\ell_1}{J_2}+\frac{4\ell_1}{J_3}+\frac{4\ell_3}{J_1}+\frac{4\ell_3}{J_2}\Big)+(\ln{J_3}+\gamma_E)\Big(\frac{1}{J_3}+\frac{4\ell_1}{J_2}+\nonumber \\
    &+\frac{4\ell_1}{J_3}+\frac{4\ell_2}{J_1}+\frac{4\ell_2}{J_3}\Big)+ S_1(\ell_1)\Big(\frac{2}{J_1} - \frac{4\ell_1}{J_2} - \frac{4 \ell_1}{J_3} - \frac{2\ell_2}{J_1} - \frac{2\ell_2}{J_3} - \frac{2\ell_3}{J_1} - \frac{2\ell_3}{J_2}\Big)+\nonumber\\ 
    &+S_1(\ell_2)\Big(\frac{2}{J_2}-\frac{4\ell_2}{J_1}-\frac{4\ell_2}{J_3}-\frac{2\ell_1}{J_2}-\frac{2\ell_1}{J_3}-\frac{2\ell_3}{J_1}-\frac{2\ell_3}{J_2}\Big)+ S_1(\ell_3)\Big(\frac{2}{J_3} - \frac{4\ell_3}{J_1}+\nonumber\\
    &- \frac{4 \ell_3}{J_1} - \frac{2\ell_1}{J_2} - \frac{2\ell_1}{J_3} - \frac{2\ell_2}{J_1} - \frac{2\ell_2}{J_3}\Big)+O(1/J_i^2)\label{Corder1}
\end{align}
where $\gamma_E$ is the Euler's constant.

Our goal now is to use the large spin expansion to write the one-loop structure constant using the basis of binomials akin to (\ref{BiBianchi}). We again divide the structure constant expression in three factors
\begin{equation}
    \hat{C}^{J_1,J_2,J_3}_{\ell_1,\ell_2,\ell_3}= X^{J_1,J_2,J_3}_{\ell_1,\ell_2,\ell_3}+Y^{J_1,J_2,J_3}_{\ell_1,\ell_2,\ell_3}+Z^{J_1,J_2,J_3}_{\ell_1,\ell_2,\ell_3}
    \label{C3spin}
\end{equation} 

The first factor corresponds to terms which in the large spin limit come multiplying logs. These terms are easier to obtain for two reasons. The first one is simply that we have more data in the large spin expansion for them ($O(1/J^5)$). The second is transcendentality: the harmonic numbers account for transcendentality one so the factors that come multiplying them turned out to be simpler than one naively would expect for the full one-loop structure constant.

Here the big advantage of the binomials representation comes to play. We can parametrize families of terms in the large spin expansion using a basis of the binomials $\mathbb{B}$. For example, when expanding in large spin we find the following factors
\begin{align}
  \hat{C}^{J_1,J_2,J_3}_{\ell_1,\ell_2,\ell_3}= \dots + 4\ln{J_1}\Bigg(\frac{\ell_2^2}{J_1J_3}-\frac{\ell_2^2}{J_1 J_3^2} - \frac{\ell2^2}{J_1^2 J_3} + \frac{\ell_1 \ell_2^2}{J_1 J_3^2} + \frac{\ell_2^3}{J_1 J_3^2} + \frac{\ell_2^3}{J_1^2 J_3}+O(J^3)\Bigg)+\dots
  \label{expampleData}
\end{align}
so we consider a linear combination of sums involving ratios of the binomials $\mathbb{B}(J_1,\ell_2)$, $\mathbb{B}(J_3,\ell_2)$, $\mathbb{B}(J_1,\ell_1)$, $\mathbb{B}(J_1-\ell_1)$, $\mathbb{B}(J_3-\ell_1,\ell_2)$ and fix their coefficients by matching with the large spin expansion (\ref{expampleData}). For this piece of the one-loop structure constant, we obtain the following expression
\begin{align}
\hat{C}^{J_1,J_2,J_3}_{\ell_1,\ell_2,\ell_3}= \dots+
    4S_1(J_1)\Bigg(\sum_{p=1}^{\infty}\frac{\mathbb{B}(J_1,\ell_2-p)}{\mathbb{B}(J_1,\ell_2)}\frac{\mathbb{B}(J_3-\ell_1,\ell_2-p)}{\mathbb{B}(J_3-\ell_1,\ell_2)}\frac{1}{p}\Bigg)+\dots
\end{align}
a simple check of this expression is to expand it in the large spin limit and recover (\ref{expampleData}). 

By following these procedure we were able to write the first factor in (\ref{C3spin}) of the one-loop structure constant, it reads
\begingroup\makeatletter\def\f@size{10}\check@mathfonts
\def\maketag@@@#1{\hbox{\m@th\large\normalfont#1}}%
\begin{align}
    X^{J_1,J_2,J_3}_{\ell_1,\ell_2,\ell_3} & = \Big(8 S_1(J_1)^2 - 4 S_1(J_1) S_1(2 J_1) + 4 S_1(J_1) S_1(J_2) + 4 S_1(J_1) S_1(J_3) +  2 S_1(\ell_1) S_1(J_1 - \ell_2)+ \nonumber\\
    & - 4 S_1(J_1) S_1(J_3 - \ell_2) - 4 S_1(J_1) S_1(\ell_2) +  2 S_1(J_1 - \ell_2) S_1(\ell_2) - S_1(\ell_2)^2 + 2 S_1(\ell_1) S_1(J_1 - \ell_3)+ \nonumber\\
    & -  4 S_1(J_1) S_1(J_2 - \ell_3) - 4 S_1(J_1) S_1(J_1 - \ell_2 - \ell_3) +  2 S_1(\ell_2) S_1(J_1 - \ell_2 - \ell_3) - 4 S_1(J_1) S_1(\ell_3)+ \nonumber\\
    & - 2 S_1(\ell_2) S_1(\ell_3) +  2 S_1(J_1 - \ell_3) S_1(\ell_3) + 2 S_1(J_1 - \ell_2 - \ell_3) S_1(\ell_3) - S_1(\ell_3)^2 -  2 S_2(J_1) - S_2(\ell_2)+ \nonumber\\
    &- S_2(\ell_3)\Big) + 4\Big(S_1(J_1)-\frac{S_1(\ell_2)}{2}-\frac{S_1(\ell_3)}{2}\Big)\Bigg(\sum_{p=1}^{\infty}\sum_{q=1}^{\infty}\frac{\mathbb{B}(J_2,\ell_3-p)}{\mathbb{B}(J_2,\ell_3)}\frac{\mathbb{B}(J_3,\ell_2-q)}{\mathbb{B}(J_3,\ell_2)}\frac{\mathbb{B}(\ell_2,p)\mathbb{B}(\ell_3,q)}{B_{(p+q)}^{(1+J_1-\ell_2-\ell_3)}}+\nonumber\\
    &+\sum_{p=1}^{\infty}\frac{\mathbb{B}(J_2,\ell_3-p)}{\mathbb{B}(J_2,\ell_3)}\frac{\mathbb{B}(J_1-\ell_2,\ell_3-p)}{\mathbb{B}(J_1-\ell_2,\ell_3)}\frac{1}{p}+\sum_{p=1}^{\infty}\frac{\mathbb{B}(J_3,\ell_2-p)}{\mathbb{B}(J_3,\ell_2)}\frac{\mathbb{B}(J_1-\ell_3,\ell_2-p)}{\mathbb{B}(J_1-\ell_3,\ell_2)}\frac{1}{p}\Bigg)+\nonumber\\
    & \left(\{J_1,\ell_1\}\longleftrightarrow\{J_2,\ell_2\} \right)+\left(\{J_1,\ell_1\}\longleftrightarrow\{J_3,\ell_3\} \right)\label{T1Terms}
\end{align}\endgroup
where $B_{(a)}^{(b)}$ is the inverse of the Euler's beta function, $B_{(a)}^{(b)}=1/B(a,b)$.

With this factors fully fixed we can turn now to the match with the Wilson loop. In order to compare with the Wilson loop we will consider the expansion around the origin limit of (\ref{defOriginLimit}). The factor (\ref{T1Terms}) written above encodes all the contributions proportional to logs, therefore it is precisely this factor that should reproduce the $\mathcal{A}_i$ factor of (\ref{repAB}). And indeed, up to order $O(1/J^{10})$ in the origin expansion we find a perfect match between the structure constant and the Wilson loop for the terms linear in logs. For finite $J_i/\ell_j$ ratio, we can perform the sums akin to (\ref{binomialsum}) and use the (\ref{lUmap}) to write the combinations of spins and polarizations in terms of cross-ratios recovering precisely that $\mathcal{A}_i=2\ln(1-U_i)$, in perfect agreement with the Wilson loop (\ref{repAB}).

The second term of (\ref{C3spin}) in the large spin limit is given only by powers of the polarizations $\ell_i$. By transcendentality the combination of binomials appearing here can be more complicated than before, as happens in the two spins case of (\ref{BiBianchi}). This expansion is again separated in various families depending in the $J_i$ and $\ell_i$ that they display, which then we try to match with a linear combination of sum of ratio of binomials. However, we were not able to fix all the combinations of $\mathbb{B}$ in a close form like (\ref{T1Terms}), these partial results we display below
\begingroup\makeatletter\def\f@size{10}\check@mathfonts
\def\maketag@@@#1{\hbox{\m@th\large\normalfont#1}}%
\begin{align}
    Y^{J_1,J_2,J_3}_{\ell_1,\ell_2,\ell_3} & = 4\Bigg(\sum_{p=1}^{\infty}\frac{\mathbb{B}(J_1,\ell_2-p)}{\mathbb{B}(J_1,\ell_2)}\frac{\mathbb{B}(J_3,\ell_2-p)}{\mathbb{B}(J_3,\ell_2)}\frac{S_1(\ell_2)-S_1(p)-S_1(\ell_2-p)}{p}-\sum_{p=1}^{\infty}\frac{\mathbb{B}(J_3,\ell_1-p)}{\mathbb{B}(J_3,\ell_1)}\frac{S_1(p-1)}{p}+\nonumber\\
    &+\sum_{p=1}^{\infty}\sum_{q=1}^{p-1}\frac{\mathbb{B}(J_1,\ell_2-p)}{\mathbb{B}(J_1,\ell_2)}\frac{\mathbb{B}(J_3,\ell_2-q)}{\mathbb{B}(J_3,\ell_2)}\frac{1}{(p-q)p}+\sum_{p=1}^{\infty}\sum_{q=p+1}^{\infty}\frac{\mathbb{B}(J_1,\ell_2-p)}{\mathbb{B}(J_1,\ell_2)}\frac{\mathbb{B}(J_3,\ell_2-q)}{\mathbb{B}(J_3,\ell_2)}\frac{1}{(q-p)q}+\nonumber\\
    &-\sum_{p=1}^{\infty}\sum_{q=1}^{\infty}\frac{\mathbb{B}(J_1,\ell_3-p)}{\mathbb{B}(J_1,\ell_3)}\frac{\mathbb{B}(J_3,\ell_1-p)}{\mathbb{B}(J_3,\ell_1)}\frac{1}{2pq}-\sum_{p=1}^{\infty}\frac{\mathbb{B}(J_1,\ell_3-p)}{\mathbb{B}(J_1,\ell_3)}\frac{S_1(p-1)}{p}\Bigg)+\nonumber \\
    &+\left(\{J_1,\ell_1\}\longleftrightarrow\{J_2,\ell_2\} \right)+\left(\{J_3,\ell_3\}\longleftrightarrow\{J_1,\ell_2\} \right)    \label{T2Terms}
\end{align}\endgroup

More precisely, we were not able to fix the combinations of binomials in a close form, for factors that in the large spin expansion mix the following spins and polarizations $\frac{\ell_{i-1}\ell_{i+1}}{J_i}$, $\frac{\ell_{i-1}\ell_{i+1}}{J_i J_{i+1}}$ and $\frac{\ell_{i-1}\ell_{i+1}}{J_{i+1}J_i J_{i+1}}$. The last term in (\ref{C3spin}) accounts for that
\begin{equation}
    Z^{J_1,J_2,J_3}_{\ell_1,\ell_2,\ell_3} = q\left(J_1,J_2,J_3,\ell_1,\ell_2\right)+q\left(J_2,J_3,J_1,\ell_2,\ell_3\right)+q\left(J_3,J_1,J_2,\ell_3,\ell_1\right)
\end{equation}
where
\begin{align}
    q\left(j_1,j_2,j_3,l_1,l_2\right) & = \sum_{n=0}^{l_2}\sum_{m=0}^{l_1}\sum_{p=0}^{l_1+l_2}\beta^{n,m,p}_{l_1,l_2}\frac{\mathbb{B}(j_1,l_2-n)}{\mathbb{B}(j_1,l_2)}\frac{\mathbb{B}(j_2,l_1-m)}{\mathbb{B}(j_2,l_1)}\frac{\mathbb{B}(j_3,l_1+l_2-p)}{\mathbb{B}(j_3,l_1+l_2)}    \label{qAnsatz2}
\end{align}
but now it has no $\gamma^{n,m,p}_{l_i,l_j}$ since the coefficients of the logs were already fixed. The remaining unknown coefficients $\beta^{n,m,p}_{l_i,l_j}$ are all fixed for $\ell_{\max}=4$ and only six remain unfixed for $\ell_{\max}=5$. These fixed $\beta^{n,m,p}_{l_i,l_j}$ are a set of simple numbers,
\begin{equation}
    \begin{array}{llllll}
        \beta_{0,0}^{0,0,0}=0, & \beta_{0,1}^{0,0,0}=0, & \beta_{0,2}^{0,0,0}=0, & \beta_{0,3}^{0,0,0}=0, & \beta_{0,3}^{0,0,0}=0,  & \beta_{0,1}^{0,0,1}=0,\vspace{0.25cm} \\ 
        \beta_{0,2}^{0,0,1}=1, & \beta_{0,3}^{0,0,1}=\frac{2}{3}, & \beta_{0,1}^{1,0,1}=0, & \beta_{0,2}^{1,0,1}=-1, & \beta_{0,3}^{1,0,1}=-\frac{2}{3}, & \dots
        \label{TabBetas}
        \end{array}
\end{equation}
which we were not able to find a pattern for. This and the other fixed coefficients are in the attached \texttt{Mathematica} notebook.

Finally, let's consider the relation with the Wilson loop. The only factor left to match in (\ref{repAB}) is $\mathcal{B}$. Since we lack a close expression for the $\beta^{n,m,p}_{l_i,l_j}$ we cannot recover the full $\mathcal{B}$ in terms of cross-ratios, however using the $\beta^{n,m,p}_{l_i,l_j}$ fixed through data we were able to match the Wilson Loop expansion up to order six, meaning we matched the first 873 terms of the origin expansion.

\end{document}